\documentclass[12pt]{article}%
\usepackage{heck}
\usepackage{cite}
\usepackage{graphicx}
\usepackage{makeidx}
\usepackage{multicol}
\usepackage{geometry}
\usepackage{amsfonts}
\usepackage{amssymb}
\usepackage{amsmath}%
\providecommand{\U}[1]{\protect\rule{.1in}{.1in}}
\begin{document}

\date{July, 2007}

\preprint{arXiv:0707.4011 \\ HUTP-07/A004 }

\institution{HarvardU}{Jefferson Physical Laboratory, Harvard University, Cambridge,
MA 02138, USA}%

\title{Geometrically Induced Phase Transitions at Large $N$}%
%

\authors{Jonathan J. Heckman\worksat{\HarvardU,}\footnote{e-mail: {\tt
jheckman@fas.harvard.edu}} and
Cumrun Vafa\worksat{\HarvardU}\footnote{e-mail: {\tt vafa@physics.harvard.edu}%
}}%

\abstract{Utilizing the large $N$ dual description of a metastable
system of branes and anti-branes wrapping rigid homologous $S^{2}$'s in a non-compact
Calabi-Yau threefold, we study phase transitions induced by changing the positions of the $S^{2}$'s.
At leading order in $1/N$ the effective potential for this system is computed by the planar limit of an auxiliary
matrix model.  Beginning at the two loop correction,
the degenerate vacuum energy density of
the discrete confining vacua split, and a potential is generated for the axion.
Changing the relative positions of the $S^{2}$'s causes discrete jumps in the energetically
preferred confining vacuum and can also obstruct direct brane/anti-brane
annihilation processes.   The branes must hop to nearby $S^{2}$'s before annihilating,
thus significantly increasing the lifetime of the corresponding non-supersymmetric
vacua.  We also speculate that misaligned metastable glueball phases may
generate a repulsive inter-brane force which stabilizes the radial mode present in compact Calabi-Yau
threefolds.}%

\maketitle

\section{Introduction}

Metastable vacua of supersymmetric string and field theories possess the
attractive feature that in contrast to a generic non-supersymmetric system,
the underlying supersymmetry of the theory often provides better control over
the dynamics of the vacuum. \ String theory realizations of metastable vacua
have been discussed in
\cite{VafaLargeN,KachruPearsonVerlinde,KKLT,KachruMcGreevy,KachruFranco,ABSV,VerlindeMonopole,HSV,KachruFrancoII,MurthyMeta,MPS}%
. \ Recent progress in finding non-supersymmetric metastable vacua in
supersymmetric QCD-like field theories was achieved in \cite{ISS} and
subsequent string theory realizations of this work
\cite{OoguriOokouchi,FrancoUranga,AhnSQCD,MQCDSeibergShih,AhnMtheory,TatarMeta,AhnOSIXPLANES,AhnMoreMeta,AhnFlavorGauged,GiveonKutasov,AhnProduct,AhnFiveNS,AhnOPLANEGivKuta,AhnMormetanoD6,TatarMetaTWO}%
.

Combining these stringy insights with the dual closed string description of
large $N$ open string systems such as the AdS/CFT correspondence
\cite{juanAdS,gkPol,witHolOne} and geometric transitions
\cite{VafaLargeN,KlebanovStrassler,MaldacenaNunezYangMills}, it was shown in
\cite{ABSV} that much of the rigid structure of $\mathcal{N}=2$ supersymmetry
remains intact for metastable vacua in type IIB\ string theory given by
D5-branes and anti-D5-branes wrapping distinct minimal size $S^{2}$'s in a
non-compact Calabi-Yau geometry of the form:%
\begin{equation}
y^{2}=W^{\prime}(x)^{2}+uv \label{rescon}%
\end{equation}
where $W^{\prime}(x)=g(x-a_{1})\cdots(x-a_{n})$ is a polynomial of degree $n$,
the variables $x,y,u,v\in%
\mathbb{C}
$ and the minimal size $S^{2}$'s are located at $x=a_{i}$. \ The $a_{i}$
correspond to non-normalizable modes in the non-compact geometry which
determine the relative separation between the branes. \ In the following we
shall denote the number of branes (resp. anti-branes) by positive (resp.
negative) integers. \ In the closed string dual description the branes (resp.
anti-branes) wrapping $S^{2}$'s are replaced by $S^{3}$'s threaded by some
amount of positive (resp. negative) RR flux. \ The geometry of the closed
string dual description is given by blowing down the $S^{2}$ and introducing a
complex deformation of equation (\ref{rescon}) by a degree $n-1$ polynomial in
$x$:%
\begin{align}
y^{2}  &  =W^{\prime}(x)^{2}+f_{n-1}(x)+uv\label{deformedcon}\\
&  =g^{2}\underset{i=1}{\overset{n}{%
{\displaystyle\prod}
}}(x-a_{i}^{+})(x-a_{i}^{-})+uv\text{.} \label{splitrootdef}%
\end{align}
This deformation splits the roots of $W^{\prime}(x)^{2}$ to $a_{i}^{+}$ and
$a_{i}^{-}$ for all $i$ and thus defines a Riemann surface with $n$ branch
cuts located near each of these roots.

At leading order in the $1/N$ expansion, the fluxes spontaneously break all of
the $\mathcal{N}=2$ supersymmetry of the Calabi-Yau compactification. \ Due to
the fact that the moduli space is still governed by rigid special geometry,
the matrix model techniques developed in
\cite{DijkgraafVafaI,DijkgraafVafaII,DijkgraafVafaIII,PerturbativeMatrixModels}
still determine all higher order corrections to the form of the periods in the
closed string dual. \ In the open string description these higher order
corrections correspond to multi loop contributions to the glueball potential.
\ In \cite{HSV} a rich phase structure for the two cut system was uncovered by
including two loop corrections to the effective potential.

\begin{figure}
[ptb]
\begin{center}
\includegraphics[
height=0.9271in,
width=3.7144in
]%
{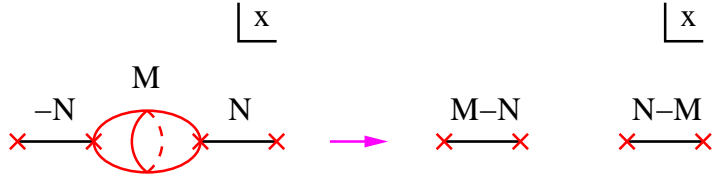}%
\caption{Depiction of flux line annihilation in a two cut geometry which
initially consists of $N>0$ units of flux through one cut and $-N$ through the
other. \ In the depiction on the left, $M>0$ D5-branes wrap the interpolating
$3$-cycle between the two $S^{3}$'s supported by flux. \ In the Minkowski
spacetime the stack of D5-branes separates a bubble of vacuum with flux
numbers $(M-N,N-M)$ from one with flux numbers $(-N,N)$. \ The end of the
annihilation process and the corresponding flux numbers are shown on the
right.}%
\label{eatflux}%
\end{center}
\end{figure}

\begin{figure}
[ptb]
\begin{center}
\includegraphics[
height=2.1932in,
width=4.1027in
]%
{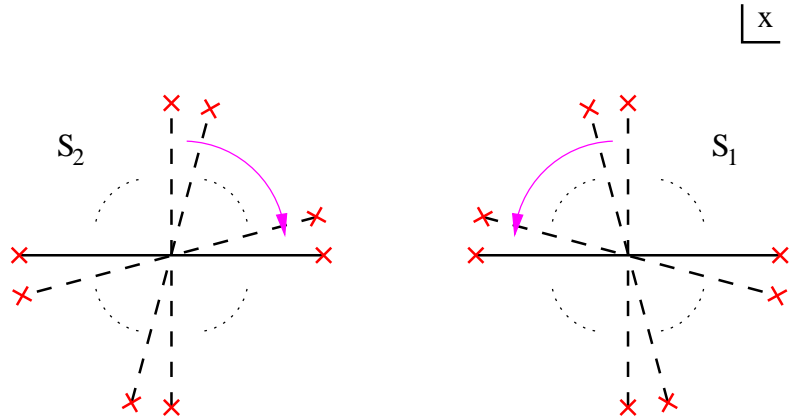}%
\caption{Two loop corrections to the glueball potential lift the degeneracy in
energy density between the confining vacua of the theory. \ These vacua
correspond to distinct orientations of the branch cuts in the closed string
dual geometry. \ The metastable branch cut orientations denoted by dashed
lines eventually decay to the energetically preferred configuration denoted by
a solid line.}%
\label{cutalign}%
\end{center}
\end{figure}

To frame some of the discussion to follow, we now review the two distinct ways
in which the two cut system exhibits metastable behavior \cite{HSV}. \ One
decay mode is given by direct flux line annihilation where the domain wall
delimiting the bubble of vacuum with lower flux is a stack of D5-branes
wrapping the compact interpolating $3$-cycle between the two $S^{3}$'s
supported by flux. \ See figure \ref{eatflux} for a depiction of this process.
\ Two loop effects reveal another way in which the vacua of the
brane/anti-brane system are metastable. \ Whereas at one loop order the
distinct confining vacua of the theory are energetically degenerate, beginning
at two loop order this degeneracy is lifted.\ \ At large $N$ and when the
scale of confinement is not exponentially suppressed, the many confining vacua
of the theory are all metastable. \ Geometrically, these vacua correspond to
distinct orientations of the branch cuts. \ The decay to the lowest energy
confining vacuum corresponds to a re-alignment of the branch cuts. \ See
figure \ref{cutalign} for a depiction of this decay process. \ \

Although the analysis of brane/anti-brane configurations with two minimal size
$S^{2}$'s should provide an adequate description of \textquotedblleft two
body\textquotedblright\ interactions, a generic compact Calabi-Yau threefold
will typically contain a large number of rigid homologous $S^{2}$'s which may
also be wrapped by branes and anti-branes. \ For example, the quintic
Calabi-Yau threefold contains $2875$ rational curves of precisely this type.
\ Perhaps surprisingly, we find that the presence of additional
\textit{unwrapped} $S^{2}$'s introduces a further layer of phase structure.

As already mentioned in the two cut case, just as a supersymmetric
configuration of $N_{i}$ branes contains $Tr(-1)^{F}=N_{1}\cdots N_{n}$
energetically degenerate confining vacua, at one loop order the vacua of the
brane/anti-brane system also remain energetically degenerate. \ It follows
from an analysis similar to the two cut case that the two loop contribution to
the energy density lifts this degeneracy. \ We also find, however, that
changing the location of the minimal size $S^{2}$'s produces discrete jumps in
the energetically preferred confining vacuum.

Interpreting the two loop energy density as a potential for the axion of our
theory, we next study the minima of this potential as a function of the
$a_{i}$'s. \ We find that the presence of additional unwrapped $S^{2}$'s
contributes a moduli space of values such that the effective $\theta$-angle of
each brane theory relaxes to zero at the minimum of the axion potential. \ We
next consider symmetric configurations where all $S^{2}$'s are occupied and
show that the discrete symmetries of the geometry translate into constraints
on strong CP\ violation in brane/anti-brane configurations which remain
invariant under the discrete symmetry $x\mapsto\overline{x}$.

Although a general analysis of the critical points of the $n$-cut system
appears quite difficult using present techniques, to keep the analysis
tractable we study configurations with $N_{1}=N$ D5-branes and $N_{2}=-N$
anti-D5-branes wrapping minimal size $S^{2}$'s in a geometry defined by a
collection of minimal size $S^{2}$'s which all lie on the real axis and remain
invariant under the map $x\mapsto-x$. \ We find that metastability is lost
once the effective 't Hooft coupling becomes too large. \ At this point, the
size of the corresponding branch cuts will begin to expand. \ A generic
configuration of additional $S^{2}$'s will typically obstruct a direct
collision between the cuts supported by positive and negative flux. \ Such
obstructions can lead to phase transitions to non-K\"{a}hler geometries of the
type found in \cite{HSV}.

The additional $S^{2}$'s can also obstruct direct brane/anti-brane
annihilation. \ Utilizing the partial classification of BPS\ states of
geometrically engineered $\mathcal{N}=2$ gauge theories obtained in
\cite{ShapereVafa}, we show that the presence of additional minimal size
$S^{2}$'s can sometimes cause the most efficient means of annihilation to
proceed via a multi-domain wall process where a brane must first tunnel to an
unoccupied minimal size $S^{2}$ before annihilating against an anti-brane.

Combining the analysis of the previous sections, we also briefly comment on
the stabilization of the radial mode which is present in compact Calabi-Yau
threefolds. \ Whereas the one loop contribution to the vacuum energy density
generates a logarithmic Coulomb attraction term, the two loop contribution
generates a power law contribution which is repulsive for appropriate glueball
phases. \ Although suggestive, we find that the value of the radial mode is
very close to the value which would destabilize the size of the glueball fields.

The rest of this paper is organized as follows. \ In section \ref{review} we
review the conjectured geometric transition for the brane/anti-brane system
and briefly discuss the phase structure found in \cite{HSV}. \ In this same
section we also include a short discussion on the regime of string theory
parameters for which we expect to have an accurate description of the low
energy dynamics. \ In section \ref{MatrixModel} we compute the two loop
contribution to the period matrix of the $n$-cut geometry. \ We briefly
discuss the form of the effective potential for geometries with unoccupied
$S^{2}$'s in section \ref{Unoccupied}. \ We next compute corrections to the
vacuum energy density as well as constraints on axion potentials in sections
\ref{twoloopen} and \ref{StrongCP}, respectively. \ In section \ref{Z2symm} we
introduce a subclass of critical points in the presence of additional $S^{2}%
$'s and study the breakdown of metastability in section \ref{Breakdown}. \ In
section \ref{DecayModes} we study brane/anti-brane annihilation processes and
in section \ref{Radion} we speculate on the stabilization of the radial mode.
\ Section \ref{Conclude} presents our conclusions.

\section{Geometrically Induced Metastability\label{review}}

In this paper we consider metastable vacua of type IIB\ string theory. \ More
precisely, we study D5-branes and anti-D5-branes which fill Minkowski space
and wrap $n$ minimal size rigid homologous $S^{2}$'s of a local Calabi-Yau
threefold defined by the hypersurface of equation (\ref{rescon}). \ This
configuration is metastable because the tension of the branes creates a
potential barrier against moving off of the minimal size $S^{2}$. \ We now
review some further properties of this system, closely following the
discussion in \cite{ABSV,HSV}. \ At each point in the complex $x$-plane the
area of the $S^{2}$ is:%
\begin{equation}
A(x)=\left(  \left\vert W^{\prime}(x)\right\vert ^{2}+\left\vert r\right\vert
^{2}\right)  ^{1/2}%
\end{equation}
where $r$ denotes the stringy volume of the minimal size $S^{2}$'s given by
turning on a non-trivial NS $B_{2}$ field. \ The bare gauge coupling of the
open string system is:%
\begin{equation}
-\alpha\left(  \Lambda_{0}\right)  =\frac{4\pi i}{g_{\mathrm{YM}}^{2}}%
+\frac{\theta_{\mathrm{YM}}}{2\pi}=\frac{4\pi i}{g_{s}}\underset{S^{2}}{\int
}B_{2}+\underset{S^{2}}{\int}C_{2}%
\end{equation}
where $C_{2}$ denotes the RR\ two form. \ In the large $N$ limit,
$N\rightarrow\infty$ but $g_{s}$ scales as $N^{-1}$ so that the associated 't
Hooft coupling $g_{s}N$ remains finite.

\begin{figure}
[ptb]
\begin{center}
\includegraphics[
height=2.0479in,
width=4.4339in
]%
{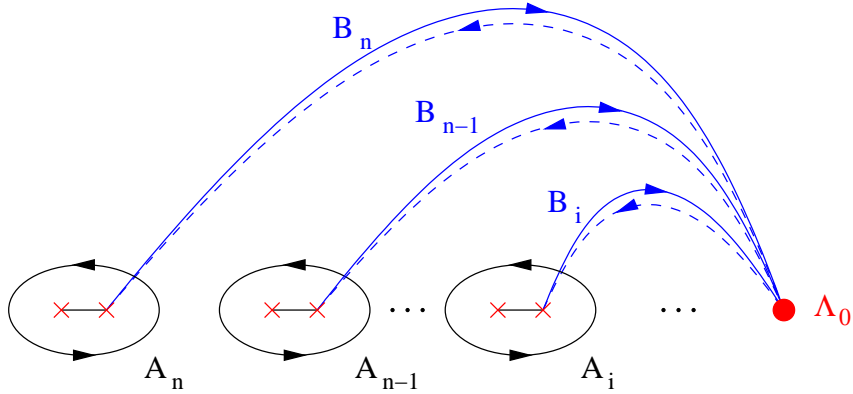}%
\caption{Depiction of the complex $x$-plane corresponding to the Riemann
surface defined by equation (\ref{deformedcon}) with $uv=0$. \ The compact
$A$-cycles reduce to counterclockwise contours which encircle each of the $n$
branch cuts of the Riemann surface. \ The non-compact $B$-cycles reduce to
contours which extend from $x=\Lambda_{0}$ on the lower sheet (dashed lines)
to $x=\Lambda_{0}$ on the upper sheet\ (solid lines).}%
\label{Riemann}%
\end{center}
\end{figure}

In the holographic dual description, the branes and anti-branes wrapping $n$
homologous $S^{2}$'s of the original geometry are replaced by fluxes threading
the $n$ topologically distinct $S^{3}$'s of the new geometry. \ The local
Calabi-Yau threefold after the transition is defined by equation
(\ref{deformedcon}) where the coefficients of $f_{n-1}(x)$ correspond to the
$n$ normalizable complex deformation parameters of the Calabi-Yau. \ This
complex equation defines a two-sheeted Riemann surface fibered over the $u$
and $v$ coordinates. \ See figure\ \ref{Riemann} for a depiction of this
geometry.

The $n$ $S^{3}$'s correspond to $n$ 3-cycles $A_{i}$ such that $A_{i}\cap
A_{j}=0$ for all $i,j$. \ Dual to each $A$-cycle is a non-compact $B$-cycle
such that $A_{i}\cap B_{j}=-B_{j}\cap A_{i}=\delta_{ij}$ and $B_{i}\cap
B_{j}=0$ for all $i,j$. \ At the level of the Riemann surface, the $A_{i}$
reduce to $n$ distinct counterclockwise contours encircling each of the $n$
branch cuts of the Riemann surface and the $B_{i}$ reduce to contours which
extend from the point $x=\Lambda_{0}$ on the lower sheet to the point
$x=\Lambda_{0}$ on the upper sheet. \ The IR\ cutoff defined by $\Lambda_{0}$
in the geometry is identified with a UV\ cutoff in the open string
description. \ The periods of the holomorphic three form $\Omega$ along the
cycles $A_{i}$ and $B_{i}$ define a basis of special coordinates for the
complex structure moduli space:%
\begin{equation}
S_{i}=\underset{A_{i}}{\int}\Omega,\text{ \ \ \ \ }\Pi_{i}=\text{\ }%
\frac{\partial\mathcal{F}_{0}}{\partial S_{i}}=\underset{B_{i}}{\int}\Omega
\end{equation}
where $\mathcal{F}_{0}$ denotes the genus zero prepotential. \ In the absence
of fluxes, each $S_{i}$ corresponds to the scalar component of a $U(1)$
$\mathcal{N}=2$ vector multiplet. \ Once branes are introduced, each $S_{i}$
is identified in the open string description with the size of a gaugino
condensate. \ Defining the period matrix:%
\begin{equation}
\tau_{ij}=\frac{\partial\Pi_{i}}{\partial S_{j}}=\frac{\partial^{2}%
\mathcal{F}_{0}}{\partial S_{i}\partial S_{j}},
\end{equation}
to leading order in the $1/N$ expansion, the K\"{a}hler metric for the
effective field theory is $\operatorname{Im}\tau_{ij}$. \ For future use we
also introduce the Yukawa couplings:%
\begin{equation}
\mathcal{F}_{ijk}\equiv\frac{\partial^{3}\mathcal{F}_{0}}{\partial
S_{i}\partial S_{j}\partial S_{k}}.
\end{equation}
\newline The net three form flux after the system undergoes a geometric
transition is:%
\begin{equation}
H_{3}=H_{\mathrm{RR}}+\tau_{\mathrm{IIB}}H_{\mathrm{NS}}%
\end{equation}
where $H_{\mathrm{RR}}$ is the net RR\ three form field strength,
$H_{\mathrm{NS}}$ is the net NS three form field strength and $\tau
_{\mathrm{IIB}}=C_{0}+ie^{-\phi}$ is the type IIB\ axio-dilaton.
\ Explicitly,
\begin{equation}
N_{i}=\underset{A_{i}}{\int}H_{3},\text{ \ \ \ \ \ }\alpha=\alpha
_{i}=-\underset{B_{i}}{\int}H_{3} \label{fluxes}%
\end{equation}
for all $i$. \ Assuming that the fluxes spontaneously break the $\mathcal{N}%
=2$ supersymmetry of the system, the flux induced effective potential is
\cite{ABSV}:%
\begin{equation}
V_{\mathrm{eff}}=\left(  \alpha_{k}+N^{k^{\prime}}\tau_{k^{\prime}k}\right)
\left(  \frac{1}{\operatorname{Im}\tau}\right)  ^{kl}\left(  \overline{\alpha
}_{l}+\overline{\tau}_{ll^{\prime}}N^{l^{\prime}}\right)  +\frac{8\pi
}{g_{\mathrm{YM}}^{2}}\left(  N_{1}+\cdots+N_{n}\right)  \label{Veffdef}%
\end{equation}
where we have introduced a constant shift in order to properly compare the
tension of branes and anti-branes \cite{ABSV}. \ Near the conifold point the
entries of the $\tau$ matrix satisfy:%
\begin{equation}
2\pi i\tau_{ij}=\delta_{ij}\log\frac{S_{i}}{W^{(2)}(a_{i})\Lambda_{0}^{2}%
}-(1-\delta_{ij})\log\frac{\Lambda_{0}^{2}}{\Delta_{ij}^{2}}+f_{ij}\left(
S_{1},...,S_{n}\right)
\end{equation}
where $\delta_{ij}$ denotes the Kronecker delta, $\Delta_{ij}=a_{i}-a_{j}$ and
$f_{ij}$ denotes an analytic power series in the variables $S_{i}$. \ The
degree $l$ terms in $f_{ij}$ correspond to the $l+1$ loop contribution to the
period matrix in the low energy effective theory.

Letting $B$ denote the set of branes and $\overline{B}$ the set of
anti-branes, the critical points of the one loop effective potential satisfy
\cite{ABSV}:%
\begin{equation}
\alpha+\underset{j\in B}{%
{\displaystyle\sum}
}\tau_{ij}N_{j}+\underset{j\in\overline{B}}{%
{\displaystyle\sum}
}\overline{\tau_{ij}}N_{j}=0 \label{oneloopvac}%
\end{equation}
\ with vacuum energy density:%
\begin{equation}
E^{(0)}=\frac{8\pi}{g_{YM}^{2}}\left(  \underset{i\in B}{\sum}\left\vert
N_{i}\right\vert +\underset{j\in\overline{B}}{\sum}\left\vert N_{j}\right\vert
\right)  -\frac{2}{\pi}\underset{i\in B,j\in\overline{B}}{\sum}\left\vert
N_{i}\right\vert \left\vert N_{j}\right\vert \log\left\vert \frac{\Lambda_{0}%
}{\Delta_{ij}}\right\vert ^{2}\text{.} \label{oneloopenergy}%
\end{equation}
Solving for the glueball fields yields:%
\begin{align}
S_{i}  &  =\zeta_{i}W^{(2)}(a_{i})\Lambda_{0}^{2}\underset{i\neq j\in B}{%
{\displaystyle\prod}
}\left(  \frac{\Lambda_{0}}{\Delta_{ij}}\right)  ^{2\left\vert \frac{N_{j}%
}{N_{i}}\right\vert }\underset{k\in\overline{B}}{%
{\displaystyle\prod}
}\left(  \frac{\overline{\Lambda_{0}}}{\overline{\Delta_{ik}}}\right)
^{2\left\vert \frac{N_{k}}{N_{i}}\right\vert }\exp\left(  -\frac{2\pi
i\alpha\left(  \Lambda_{0}\right)  }{\left\vert N_{i}\right\vert }\right)
\text{ \ \ \ (}N_{i}>0\text{)}\label{BCRIT}\\
S_{i}  &  =\zeta_{i}W^{(2)}(a_{i})\Lambda_{0}^{2}\underset{i\neq k\in
\overline{B}}{%
{\displaystyle\prod}
}\left(  \frac{\Lambda_{0}}{\Delta_{ik}}\right)  ^{2\left\vert \frac{N_{k}%
}{N_{i}}\right\vert }\underset{j\in B}{%
{\displaystyle\prod}
}\left(  \frac{\overline{\Lambda_{0}}}{\overline{\Delta_{ij}}}\right)
^{2\left\vert \frac{N_{j}}{N_{i}}\right\vert }\exp\left(  \frac{2\pi
i\overline{\alpha}\left(  \Lambda_{0}\right)  }{\left\vert N_{i}\right\vert
}\right)  \text{ \ \ \ \ \ (}N_{i}<0\text{)} \label{BBCRIT}%
\end{align}
where $\zeta_{i}$ denotes an $\left\vert N_{i}\right\vert ^{\text{th}}$ root
of unity. \ These discrete phase choices label the distinct confining vacua of
the theory.

As explained in greater detail in \cite{ABSV}, the brane/anti-brane system
decays to the supersymmetric ground state by nucleating a bubble of lower
vacuum energy density. \ The corresponding domain wall solution is given by a
D5-brane wrapping a supersymmetric 3-chain in the Calabi-Yau threefold. \ From
the perspective of the closed string dual, the D5-brane wrapping the
corresponding supersymmetric 3-cycle separates vacua with different amounts of
flux quanta.

The type IIA\ T-dual description of the brane/anti-brane system with two
minimal size $S^{2}$'s was first commented on in \cite{TatarMeta} and studied
in detail in \cite{MPS}. \ This corresponds to a configuration with D4-branes
and anti-D4-branes suspended between a pair of NS5-branes defined by the
equations $y=\pm W^{\prime}(x)$. \ The analogue of the IIB\ closed string
description maps to the regime where the D4-branes and anti-D4-branes have
dissolved as flux in the NS5-branes. \ This configuration was also lifted to
an M-theory description given by an M5-brane wrapping a harmonic rather than
holomorphic curve. \ As noted in \cite{MPS}, although the absence of
supersymmetry would seemingly preclude any direct connection between the
effective potentials of the IIB\ and IIA systems due to the difference in the
size of the T-dualized circle, the observed match in the limit $g_{s}%
\rightarrow0$ between the two effective potentials provides further evidence
that at leading order in the $1/N$ expansion supersymmetry is indeed
spontaneously broken by the presence of fluxes.

Due to the fact that a convenient parametrization of the associated M-theory
curve is not known for $n$-cut geometries, we will unfortunately not be able
to exploit the powerful techniques developed in \cite{MPS} to describe the
$g_{s}\rightarrow0$ limit of the all loop phase structure of the associated
system for even a restricted set of fluxes.

Before proceeding to a short review of the two cut phase structure, we first
comment on the possibility of additional corrections to the K\"{a}hler
potential due to $\alpha^{\prime}$ corrections. \ This is a subtle point
because in order to keep the string modes which mediate supersymmetry breaking
from decoupling we must keep $\alpha^{\prime}$ finite. \ Although in the
strict infinite volume limit such effects will apparently not alter the form
of the K\"{a}hler potential, for finite K\"{a}hler modulus such corrections
will introduce important corrections \cite{DouglasAlphaPrime}. \ Because we
work in the context of non-compact Calabi-Yau threefolds, we shall assume for
the purposes of this paper that such corrections are negligible.

\subsection{Two Cut Phase Structure}

We now review the phase structure of the two cut system \cite{HSV}. \ At
leading order in an expansion of the periods, the two$\ $cut period matrix is
\cite{CIV}:%
\begin{align}
2\pi i\tau &  =\left(
\begin{array}
[c]{cc}%
\log\frac{S_{1}}{g\Delta_{12}\Lambda_{0}^{2}} & -\log\frac{\Lambda_{0}^{2}%
}{\Delta_{12}^{2}}\\
-\log\frac{\Lambda_{0}^{2}}{\Delta_{12}^{2}} & \log\frac{S_{2}}{-g\Delta
_{12}\Lambda_{0}^{2}}%
\end{array}
\right) \\
&  +\left(
\begin{array}
[c]{cc}%
4\frac{S_{1}}{g\Delta_{12}^{3}}-10\frac{S_{2}}{g\Delta_{12}^{3}} &
-10\frac{S_{1}}{g\Delta_{12}^{3}}+10\frac{S_{2}}{g\Delta_{12}^{3}}\\
-10\frac{S_{1}}{g\Delta_{12}^{3}}+10\frac{S_{2}}{g\Delta_{12}^{3}} &
-4\frac{S_{2}}{g\Delta_{12}^{3}}+10\frac{S_{1}}{g\Delta_{12}^{3}}%
\end{array}
\right)  +\mathcal{O}\left(  S^{2}\right)  \text{.}%
\end{align}
As shown in \cite{HSV}, there is an intricate phase structure beginning at two
loop order. \ Just as a supersymmetric configuration of $N_{i}$ branes
contains $Tr(-1)^{F}=N_{1}N_{2}$ energetically degenerate confining vacua, at
one loop order the confining vacua of the brane/anti-brane system also remain
energetically degenerate. \ These vacua correspond to the different roots of
unity appearing in the one loop critical points of equations (\ref{BCRIT}) and
(\ref{BBCRIT}). \ Beginning at two loop order, this degeneracy in the energy
densities is lifted. \ In the closed string dual, the confining vacuum of
lowest energy corresponds to a configuration where the branch cuts align along
a common axis in the complex $x$-plane. \ When the size of the confinement
scale is fixed to a finite value, the action for quantum tunneling to the
energetically preferred confining vacuum scales as a positive power of $N$ so
that the corresponding tunneling rate is exponentially suppressed. \ The two
loop contribution to the effective potential also destabilizes the vacuum once
the flux becomes comparable to the value:%
\begin{equation}
g_{\mathrm{YM}}^{2}N\sim\frac{1}{\log\left\vert \frac{\Lambda_{0}}{\Delta
_{12}}\right\vert }\text{.} \label{breakdown}%
\end{equation}
Once metastability is lost, the branch cuts begin to expand until they nearly
collide. \ When this occurs the interpolating 3-cycle $B_{1}-B_{2}$ collapses
to nearly zero size. \ The contribution due to additional light states can
potentially drive the system to two qualitatively different endpoints.
\ Whereas D5-branes wrapping $B_{1}-B_{2}$ will tend to lower the flux of the
corresponding configuration so that the system relaxes back to another
metastable configuration, the contribution due to a D3-brane wrapping the same
collapsing cycle can trigger a transition to a non-K\"{a}hler geometry
\cite{HSV}.

\section{Matrix Model Computation \label{MatrixModel}}

In order to study the phase structure of more general geometries, we first
determine the two loop contribution to the period matrix in the $n$-cut
geometry. \ Although it is in principle possible to determine the form of this
correction by directly evaluating the period integrals of the closed string
dual, we shall instead use the matrix model technology developed in
\cite{DijkgraafVafaI,DijkgraafVafaII,DijkgraafVafaIII} to reduce the
computation to a perturbative Feynman diagram analysis. \ This will have the
added benefit that we will be able to isolate individual string exchange
processes which contribute to the vacuum energy density.

The genus zero prepotential of the closed string dual geometry is computed to
all orders by the planar limit of a large $N$ auxiliary matrix model with
partition function:%
\begin{equation}
Z_{MM}=\frac{1}{\text{Vol}\left(  U\left(  N\right)  \right)  }\int d\Phi
\exp\left(  -\frac{1}{g_{s}}TrW\left(  \Phi\right)  \right)
\end{equation}
where $\Phi$ is a holomorphic $N\times N$ matrix and the above matrix integral
should be understood as a contour integral. \ The prepotential of the $n$-cut
geometry near the semi-classical expansion point is given by expanding the
eigenvalues of $\Phi$ about the $n$ critical points of the polynomial $W$.
\ The usual eigenvalue repulsion term of the matrix model causes these
eigenvalues to fill the $n$ cuts of the geometry after the geometric
transition. \ With $M_{i}$ eigenvalues sitting at the $i^{th}$ cut of the
geometry, this matrix model may be recast as an $n$-matrix model of the form:%
\begin{equation}
Z_{MM}=\frac{1}{\underset{i=1}{\overset{n}{%
{\displaystyle\prod}
}}\text{Vol}\left(  U\left(  M_{i}\right)  \right)  }\int d\Phi_{11}\cdots
d\Phi_{nn}\exp\left(  \underset{i=1}{\overset{n}{\sum}}-\frac{1}{g_{s}}%
W_{i}\left(  \Phi_{ii}\right)  -\frac{1}{g_{s}}W_{int}\left(  \Phi_{11}%
,\cdots,\Phi_{nn}\right)  \right)
\end{equation}
where $\Phi_{ii}$ denotes the $M_{i}\times M_{i}$ block of $\Phi$ along its
diagonal. \ The periods of the $A$-cycles are given by the partial 't Hooft
couplings of the matrix model:%
\begin{equation}
S_{i}=g_{s}M_{i}\text{.}%
\end{equation}
Evaluating $Z_{MM}$ in the saddle point approximation, the genus zero
prepotential is:%
\begin{equation}
\mathcal{F}_{0}=\mathcal{F}_{\mathrm{non-pert}}+\mathcal{F}_{\mathrm{pert}}%
\end{equation}
where $\mathcal{F}_{\mathrm{non-pert}}$ corresponds to contributions from the
Vol$\left(  U\left(  M_{i}\right)  \right)  $ factors and $\mathcal{F}%
_{\mathrm{pert}}$ corresponds to the perturbative contributions from planar
Feynman diagrams:%
\begin{align}
2\pi i\mathcal{F}_{\mathrm{non-pert}}  &  =\underset{i=1}{\overset{n}{\sum}%
}\frac{1}{2}S_{i}^{2}\log\frac{S_{i}}{\Lambda_{0}^{3}}\\
2\pi i\mathcal{F}_{\mathrm{pert}}  &  =\underset{i=1}{\overset{n}{\sum}}%
-S_{i}W(a_{i})+\underset{0\leq i_{1},...,i_{n}}{\sum}C_{i_{1}...i_{n}}%
S_{1}^{i_{1}}\cdots S_{n}^{i_{n}}%
\end{align}
where the $C_{i_{1}...i_{n}}$ are coefficients which are in principle calculable.

We now compute all contributions to $\mathcal{F}_{\mathrm{pert}}$ proportional
to $M_{i}M_{j}M_{k}$ for all $i,j,k$. \ Following
\cite{PerturbativeMatrixModels}, the gauge-fixed matrix model action is given
by the sum of two contributions:%
\begin{equation}
S_{\mathrm{MM}}=S_{\Phi}+S_{\mathrm{ghost}}%
\end{equation}
where:%
\begin{equation}
S_{\Phi}=\frac{1}{g_{s}}\underset{r\geq0}{%
{\displaystyle\sum}
}\underset{i}{\sum}\frac{W^{(r)}\left(  a_{i}\right)  }{r!}Tr\left(  \Phi
_{ii}\right)  ^{r}%
\end{equation}
and:%
\begin{equation}
S_{\mathrm{ghost}}=\frac{1}{g_{s}}\underset{i\neq j}{\sum}Tr\left(
\Delta_{ji}B_{ij}C_{ji}+B_{ij}\Phi_{jj}C_{ji}+C_{ij}\Phi_{jj}B_{ji}\right)
\end{equation}
where $C_{ij}$ denotes a scalar ghost and $B_{ij}$ its conjugate. \ In the
above we have also introduced a constant shift in the definition of the
$\Phi_{ii}$'s. \ Suppressing all matrix indices, the propagators of the fields
are:%
\begin{align}
\left\langle \Phi_{ii}\Phi_{ii}\right\rangle  &  =\frac{g_{s}}{W^{(2)}\left(
a_{i}\right)  }\\
\left\langle B_{ij}C_{ji}\right\rangle  &  =\frac{g_{s}}{\Delta_{ij}}%
\end{align}
for all $i\neq j$. \ The $l$-point interaction vertex of $l$ $\Phi_{ii}$
fields has weight $-W^{(l)}\left(  a_{i}\right)  /g_{s}(l-1)!$ and the three
point interaction vertex between the $B$, $C$ and $\Phi$ fields has weight
$-1/g_{s}$.

\begin{figure}
[ptb]
\begin{center}
\includegraphics[
height=2.2502in,
width=4.088in
]%
{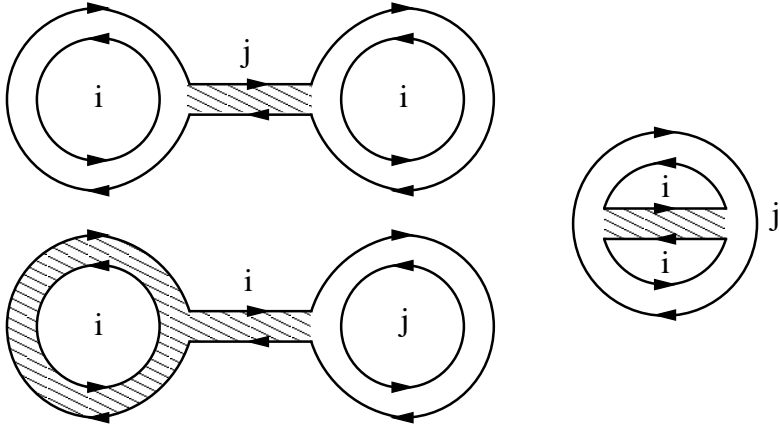}%
\caption{Depiction of all Feynman diagrams which contribute to the $M_{i}%
^{2}M_{j}$ term of $\mathcal{F}_{pert}$ for $i\neq j$. \ In each diagram,
ghost propagators are denoted by white and $\Phi$ propagators by dashed lines.
\ The index loops of each diagram are also shown.}%
\label{twoloop}%
\end{center}
\end{figure}

For $i\neq j$, the three Feynman diagrams which contribute to the $M_{i}%
^{2}M_{j}$ term of the free energy are depicted in figure \ref{twoloop}.
Proceeding from the right-most diagram to the upper left diagram in a
clockwise direction, the contribution of each diagram to the matrix model free
energy is:%
\begin{equation}
A_{iij}=-\frac{g_{s}M_{i}^{2}M_{j}}{W^{(2)}\left(  a_{i}\right)  \Delta
_{ij}^{2}}-\frac{g_{s}M_{i}^{2}M_{j}W^{(3)}\left(  a_{i}\right)  }%
{W^{(2)}\left(  a_{i}\right)  ^{2}\Delta_{ij}}+\frac{2g_{s}M_{i}^{2}M_{j}%
}{W^{(2)}\left(  a_{j}\right)  \Delta_{ij}^{2}}.
\end{equation}

For $i\neq j\neq k\neq i$, the diagrams which contribute to the $M_{i}%
M_{j}M_{k}$ term of the free energy possess the same ghost and $\Phi$ field
content as the upper left diagram of figure \ref{twoloop} but with all three
index loops distinct. \ The net contribution to the matrix model free energy
from this class of diagrams is:%
\begin{equation}
A_{ijk}=\frac{4g_{s}M_{i}M_{j}M_{k}}{W^{(2)}(a_{i})\Delta_{ij}\Delta_{ik}%
}+(i\leftrightarrow j)+(i\leftrightarrow k)\text{.}%
\end{equation}

Two of the three diagrams which contribute to the $M_{i}^{3}$ term of the free
energy are given by replacing the ghost content of the diagrams of figure
\ref{twoloop} by $\Phi$ fields and interaction vertices. \ When three or more
cuts are present, a planar figure eight diagram also contributes. \ Starting
from the right-most diagram of figure \ref{twoloop} and ending with the figure
eight diagram, the net contribution to the matrix model free energy from this
class of diagrams is:%
\begin{equation}
A_{iii}=\frac{g_{s}M_{i}^{3}W^{(3)}\left(  a_{i}\right)  ^{2}}{24W^{(2)}%
(a_{i})^{3}}+\frac{g_{s}M_{i}^{3}W^{(3)}\left(  a_{i}\right)  ^{2}}%
{8W^{(2)}(a_{i})^{3}}-\frac{g_{s}M_{i}^{3}W^{(4)}(a_{i})}{12W^{(2)}(a_{i}%
)^{2}}\text{.}%
\end{equation}
The two loop corrections to the period matrix are summarized in appendix A.

\section{Effective Potentials and Unoccupied $S^{2}$'s \label{Unoccupied}}

In this brief section we explain how to treat the effective potential for
brane/anti-brane systems with unoccupied minimal size $S^{2}$'s. \ To this
end, we first review the treatment of similar supersymmetric brane
configurations developed in \cite{ShigemoriVafa}. \ Due to the fact that all
of the initial $S^{2}$'s in the open string description are homologous, even
an unwrapped $S^{2}$ will na\"{\i}vely undergo a geometric transition to an
$S^{3}$. \ Note, however, that because there is no flux to support the $S^{3}$
counterparts of the unwrapped $S^{2}$'s, additional light degrees of freedom
corresponding to D3-branes wrapping the collapsing $S^{3}$'s will enter the
low energy dynamics. \ The effective superpotential now includes contributions
of the form:%
\begin{equation}
\alpha S+\sqrt{2}Q_{L}Q_{R}S\subset\mathcal{W}_{eff}%
\end{equation}
where $Q_{L}$ and $Q_{R}$ denote two $\mathcal{N}=1$ chiral multiplets which
combine to form the $\mathcal{N}=2$ hypermultiplet describing the D3-brane.
\ When the $Q$ fields condense, the original $S^{3}$ disappears and is
replaced by an $S^{2}$ of size $\left\langle Q_{L}Q_{R}\right\rangle
=-\alpha/\sqrt{2}$. \ The glueball superpotential for such a system is
computed by a matrix model with action given by $W(x)$ where the $S_{i}$'s
corresponding to zero flux are set to zero in the prepotential.

In the non-supersymmetric brane/anti-brane configurations studied in this
paper the effective potential is controlled by the period matrix $\tau_{ij}$.
\ Due to the fact that $\mathcal{N}=2$ supersymmetry is spontaneously broken
at leading order in $1/N$ by the presence of positive and negative flux, the
prescription for supersymmetric configurations described above will continue
to hold in the non-supersymmetric context. \ Labelling the fluxes so that
$N_{1},...,N_{l}\neq0$ for some positive integer $l\leq n$ with all other
$N_{i}$ equal to zero, the appropriate period matrix to use in equation
(\ref{Veffdef}) is given by the $l\times l$ truncation of $\tau_{ij}$ to
$1\leq i,j\leq l$ where all glueball fields corresponding to zero flux have
been formally set to zero in the prepotential.

\section{Two Loop Corrections to the Energy Density \label{twoloopen}}

At one loop order, equation (\ref{oneloopenergy}) implies that the discrete
phase choices for the glueball fields present in equations (\ref{BCRIT}) and
(\ref{BBCRIT}) have identical vacuum energy densities. \ As in the two cut
case, we find that two loop effects lift this degeneracy. \ The decay to the
confining vacuum of lowest energy proceeds via D5-branes wrapping the
$A$-cycles in the closed string dual geometry. \ We refer the reader to
\cite{HSV} for further discussion on the alignment of glueball phases due to
tunneling processes. \ In this section we study the orientation of the branch
cuts in the lowest energy confining vacuum as a function of the roots of
$W^{\prime}(x)$.

In the remainder of this paper we let $\delta\tau$ denote the two loop
correction to the period matrix. \ Expanding $V_{\mathrm{eff}}$ to linear
order in $\delta\tau$ and applying equation (\ref{oneloopvac}) yields the
first order correction to the vacuum energy density due to two loop effects:%
\begin{equation}
E=E^{(0)}+\underset{j\in B,k\in\overline{B}}{%
{\displaystyle\sum}
}4\left\vert N_{j}\right\vert \left\vert N_{k}\right\vert \operatorname{Re}%
\left(  i\delta\tau_{jk}\right)  +\mathcal{O}(\delta\tau^{2})
\label{twoloopshift}%
\end{equation}
where $\delta\tau_{jk}$ is evaluated at the one loop corrected value of the
$S_{i}$'s given by equations (\ref{BCRIT}) and (\ref{BBCRIT}). \ In the
following we shall refer to the second term as $\delta E$. \ Because the
components of $\delta\tau_{jk}$ are linear in the $S_{i}$'s, we conclude that
the two loop contribution lifts the degeneracy in energy present at one loop
order. \ Note in particular that the energy only receives contributions from
off-diagonal brane/anti-brane components of $\delta\tau_{jk}$.

From the perspective of the closed string dual geometry, the phases of the
$S_{i}$'s correspond to the direction of alignment for the branch cuts in the
complex $x$-plane. \ Expanding the defining period integrals for the $S_{i}$
in terms of $\left(  a_{i}^{+}-a_{i}^{-}\right)  $ yields:%
\begin{equation}
S_{i}=\frac{g}{2\pi i}\underset{a_{i}^{-}}{\overset{a_{i}^{+}}{\int}}%
\sqrt{\underset{i=1}{\overset{n}{%
{\displaystyle\prod}
}}(x-a_{i}^{+})(x-a_{i}^{-})}dx=\frac{1}{4}W^{(2)}(a_{i})\left(  \delta
x_{i}\right)  ^{2}+\mathcal{O}(\delta x^{3}) \label{cutsquareddef}%
\end{equation}
where $4\left(  \delta x_{i}\right)  ^{2}=(a_{i}^{+}-a_{i}^{-})^{2}$.

Although there is an ambiguity in the sign of the above period integral due to
the presence of the square root, the different choices of signs correspond to
distinct orientations of the $A$- and $B$- cycles in the Calabi-Yau threefold.
\ Indeed, once the orientation of the $B$-cycles is fixed, the choice of signs
in the above period integrals are then completely fixed. \ To determine the
appropriate sign, we consider a geometry given by branch cuts which are all
located on the real axis of the complex $x$-plane with $g,\Lambda_{0}>0$.
\ Ordering the points so that $a_{1}^{+}>a_{1}^{-}>\cdots>a_{n}^{+}>a_{n}^{-}%
$, we have:%
\begin{equation}
\Pi_{1}=\frac{g}{2\pi i}\underset{a_{1}^{+}}{\overset{\Lambda_{0}}{%
{\displaystyle\int}
}}\sqrt{\underset{i=1}{\overset{n}{%
{\displaystyle\prod}
}}(x-a_{i}^{+})(x-a_{i}^{-})}dx=\frac{R}{2\pi i}%
\end{equation}
where $R$ is a real number. \ Adhering to the sign conventions used in
\cite{CIV}, general monodromy arguments yield:%
\begin{equation}
\Pi_{1}\sim\frac{S_{1}}{2\pi i}\log\frac{S_{1}}{W^{(2)}(a_{1})\Lambda_{0}^{2}%
}+\mathcal{O}(S^{0})\text{.}%
\end{equation}
We therefore conclude that the proper choice of sign for $S_{1}$ and therefore
all of the $S_{i}$'s is the one given by equation (\ref{cutsquareddef}).

In the rest of this section we study the alignment of branch cuts in the
lowest energy confining vacuum. \ Because there are a discrete number of
confining vacua, a generic change in the parameters $a_{i}$ will typically
induce a small jump of $\mathcal{O}(1/N)$ in the preferred confining vacuum.
\ At large $N$ such small jumps do not produce a dramatic change in the
physics because we already approximate the discrete phase choice by a
continuous variable. \ Even in this limit, however, we find that an
appropriate variation of parameters can induce large discrete jumps in the
preferred confining vacuum. \ In subsection \ref{ALIGNTHREE} we treat
geometries with three minimal size $S^{2}$'s and present two examples where
the alignment of cuts naturally generalizes the results found for two cut
geometries. \ We next demonstrate that for more general configurations,
changing the relative positions of the $a_{i}$'s can induce discrete jumps in
the preferred alignment direction of the branch cuts. \ In subsection
\ref{ALIGNMULTIPLE} we generalize this analysis to configurations with
additional minimal size $S^{2}$'s.

\subsection{Alignment with Three Minimal $S^{2}$'s \label{ALIGNTHREE}}

We now study the energetics of branch cut alignment in geometries with:%
\begin{equation}
W^{\prime}(x)=g(x-a_{1})(x-a_{2})(x-a_{3})\text{.}%
\end{equation}
Our expectation is that configurations with a high degree of symmetry will
exhibit behavior which is similar to that of the two cut system.

As a first example, we take the $a_{i}$'s to form an equilateral triangle with
$N_{1}>0$ D5-branes at $x=a_{1}=a$, $N_{2}>0$ D5-branes at $x=a_{2}=-a$, and
$N_{3}<0$ anti-D5-branes at $x=a_{3}=i\sqrt{3}a$. \ Setting $\delta
x_{i}=r_{i}e^{i\beta_{i}}$ with $r_{i}>0$, the two loop contribution to the
energy is\ given by the second term of equation (\ref{twoloopshift}):%
\begin{equation}
\delta E=\operatorname{Re}\left[
\begin{array}
[c]{c}%
\frac{\left\vert N_{1}\right\vert \left\vert N_{3}\right\vert }{8\pi a^{2}%
}\left(  -1-7i\sqrt{3}\right)  r_{1}^{2}e^{2i\beta_{1}}+\frac{\left\vert
N_{2}\right\vert \left\vert N_{3}\right\vert }{8\pi a^{2}}\left(
-1+7i\sqrt{3}\right)  r_{2}^{2}e^{2i\beta_{2}}\\
+\left(  \frac{\left\vert N_{1}\right\vert \left\vert N_{3}\right\vert }{8\pi
a^{2}}\left(  11-3i\sqrt{3}\right)  +\frac{\left\vert N_{2}\right\vert
\left\vert N_{3}\right\vert }{8\pi a^{2}}\left(  11+3i\sqrt{3}\right)
\right)  r_{3}^{2}e^{2i\beta_{3}}%
\end{array}
\right]  \text{.} \label{energyequilateral}%
\end{equation}
Minimizing $\delta E$ with respect to the $\beta_{i}$'s, it now follows that
in the lowest energy configuration the cuts always point towards the interior
of the triangle. \ Bisecting each $60^{\circ}$ angle of the equilateral
triangle, we further find that the cuts threaded by positive flux align on the
side of the bisection closest to the cut with negative flux. \ Finally, it
follows from the last line of equation (\ref{energyequilateral}) that the
orientation of the cut with negative flux depends on the relative magnitudes
of $\left\vert N_{1}\right\vert $ and $\left\vert N_{2}\right\vert $.

In fact, the position of minimal size $S^{2}$'s not wrapped by branes will
also influence the alignment of the cuts in the lowest energy confining
vacuum. \ To this end, consider a configuration with $N_{1}>0$ D5-branes
located at $x=a_{1}=a>0$, $N_{2}<0$ anti-D5-branes located at $x=a_{2}=-a$ and
with a vanishingly small number of branes or anti-branes wrapping the $S^{2}$
at $x=a_{3}=c$. \ In this case $\left(  \delta x_{3}\right)  ^{2}=0$ and the
correction $\delta E$ takes the form:%
\begin{equation}
\delta E=\frac{\left\vert N_{1}\right\vert \left\vert N_{2}\right\vert }{4\pi
a^{2}}\operatorname{Re}\left(  \left(  -5-\frac{8ac}{a^{2}-c^{2}}\right)
\left(  \delta x_{1}\right)  ^{2}+\left(  -5+\frac{8ac}{a^{2}-c^{2}}\right)
\left(  \delta x_{2}\right)  ^{2}\right)  \text{.} \label{specialthreecut}%
\end{equation}
Note that in the limit $c\rightarrow\infty$,%
\begin{equation}
\underset{c\rightarrow\infty}{\lim}\delta E=-\frac{5\left\vert N_{1}%
\right\vert \left\vert N_{2}\right\vert }{4\pi a^{2}}\operatorname{Re}\left(
\left(  \delta x_{1}\right)  ^{2}+\left(  \delta x_{2}\right)  ^{2}\right)
\end{equation}
the corresponding $S^{2}$ at $x=c$ decouples from the dynamics of the theory
and we find that just as in the two cut geometry, the branch cuts align with
the real axis of the complex $x$-plane in the minimal energy configuration.

The symmetries of such configurations still constrain the direction of
alignment. \ Indeed, when the $a_{i}$'s form an isosceles triangle with $c=iL$
a pure imaginary number, we find:%
\begin{equation}
\delta E=\frac{\left\vert N_{1}\right\vert \left\vert N_{2}\right\vert }{\pi
}\operatorname{Re}\left(  Qr_{1}^{2}e^{2i\beta_{1}}+\overline{Q}r_{2}%
^{2}e^{2i\beta_{2}}\right)
\end{equation}
where $\delta x_{i}=r_{i}e^{i\beta_{i}}$ and $Q$ is a complex number. \ It
thus follows that the system has lowest energy when $\beta_{1}=-\beta_{2}$ so
that the branch cuts point symmetrically towards $x=c$. \ Indeed, expanding
equation (\ref{specialthreecut}) for small $L/a$ yields:%
\begin{equation}
\delta E=-\frac{5\left\vert N_{1}\right\vert \left\vert N_{2}\right\vert
}{4\pi a^{2}}\operatorname{Re}\left(  e^{8iL/5a}r_{1}^{2}e^{2i\beta_{1}%
}+e^{-8iL/5a}r_{2}^{2}e^{2i\beta_{2}}\right)  +\mathcal{O}\left(  \frac{L^{2}%
}{a^{2}}\right)
\end{equation}
so that both cuts tip towards $x=c$ with angle $4L/5a$. \ Although the cuts
tip so as to touch one another, they do not align at the proper angle to touch
the point $x=c$. \ This indicates that when all three cuts are of finite size,
the energetically preferred configuration corresponds to the case where the
endpoints of the cuts are maximally close to touching.%
\begin{figure}
[ptb]
\begin{center}
\includegraphics[
height=1.1243in,
width=1.7884in
]%
{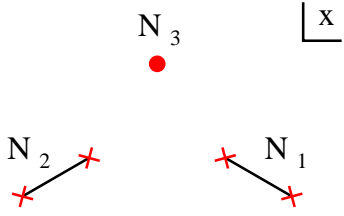}%
\caption{Depiction of branch cut orientation in the lowest energy confining
vacuum of the three cut system with $N_{1}>0$ units of flux through the cut
near $x=a$, $N_{2}<0$ units of flux through the cut near $x=-a$ and $N_{3}=0$
units of flux through the cut near $x=iL$ with $L,a>0$. \ The presence of the
additional minimal size $S^{2}$ at $x=iL$ causes the two cuts supported by
flux to tip up.}%
\end{center}
\end{figure}

The above examples suggest that although the alignment of the branch cuts in
the minimal energy confining vacuum depends on the roots of $W^{\prime}(x)$,
up to small discrete jumps of order $1/N$, this alignment is a smooth function
of the relative position of the $a_{i}$'s. \ Indeed, the coefficients
multiplying $\left(  \delta x_{i}\right)  ^{2}$ are rational functions of the
$a_{i}$. \ Note, however, that varying the $a_{i}$ will produce a discrete
jump in the alignment of the branch cuts whenever crossing through the zero
set of this rational function. \ When the position of a single root of $a_{i}$
is varied, this zero set reduces to a finite number of points in the complex
$x$-plane. \ In this case a jump can be avoided by traversing a contour which
does not pass through such a root.

To establish the existence of such jumping phenomena, we restrict to
configurations where all three $a_{i}$'s lie on the real axis of the complex
$x$-plane and $N_{1}>0>N_{2}$, and $N_{3}$ is vanishingly small. \ With
notation as in the previous example, the branch cuts in the minimal energy
confining vacuum align parallel to the real axis when $c$ lies within the
union of the following intervals:%
\begin{align}
c  &  \in\left(  -\infty,-a\right)  \cup\left(  \frac{a}{5}\left(  4-\sqrt
{41}\right)  ,a\right)  \cup\left(  \frac{a}{5}\left(  4+\sqrt{41}\right)
,\infty\right)  &  &  \Rightarrow\delta x_{1}\in%
\mathbb{R}%
\\
c  &  \in\left(  -\infty,-\frac{a}{5}\left(  4+\sqrt{41}\right)  \right)
\cup\left(  -a,\frac{a}{5}\left(  -4+\sqrt{41}\right)  \right)  \cup\left(
a,\infty\right)  \text{ } &  &  \Rightarrow\delta x_{2}\in%
\mathbb{R}%
\end{align}
and perpendicular to the real axis for all other values of $c\in%
\mathbb{R}
$ which do not cause the coefficients of $\left(  \delta x_{i}\right)  ^{2}$
to vanish. \ Hence, there exist regions on the real line where both cuts are
parallel, both are perpendicular, and one is parallel while the other is
perpendicular to the real axis of the complex $x$-plane. \ Due to the fact
that for $g>0$, $W^{(2)}(a)$ is positive for $c<a$ and negative for $c>a$ we
conclude that the glueball phases are also sensitive to the location of the
unoccupied $S^{2}$.

Although the locations in the complex $x$-plane where cut alignment phase
transitions occur appear to have little geometric content, the individual
string exchange diagrams determined by the matrix model reflect the underlying
location of the D5-branes and anti-D5-branes in the open string description.
\ Without loss of generality, we focus on the contributions to the alignment
of the cut near $x=a_{1}=a$. \ Proceeding from the right-most diagram to the
upper lefthand diagram in a clockwise direction in figure \ref{twoloop}, the
sign of the contribution of each diagram to the alignment of $\delta x_{1}$
along the real axis of the complex $x$-plane is:%
\begin{align}
sign\left(  -\frac{1}{W^{(2)}\left(  a_{1}\right)  }\frac{1}{\Delta_{12}^{2}%
}W^{(2)}(a_{1})\right)   &  \Rightarrow\delta x_{1}\in%
\mathbb{R}
\text{ \ \ \ \ \ for }c\in%
\mathbb{R}%
\\
sign\left(  -\frac{W^{(3)}\left(  a_{1}\right)  }{W^{(2)}\left(  a_{1}\right)
^{2}}\frac{1}{\Delta_{12}}W^{(2)}(a_{1})\right)   &  \Rightarrow\delta
x_{1}\in%
\mathbb{R}
\text{ \ \ \ \ \ for }c\notin\left[  a_{1},a_{1}+\Delta_{12}\right] \\
sign\left(  \frac{2}{W^{(2)}\left(  a_{2}\right)  }\frac{1}{\Delta_{12}^{2}%
}W^{(2)}(a_{1})\right)   &  \Rightarrow\delta x_{1}\in%
\mathbb{R}
\text{ \ \ \ \ \ for }c\notin\left[  a_{2},a_{1}\right]  \text{.}%
\end{align}
Note in particular that for the second $1-2$ exchange diagram the alignment of
the cut depends on the location of an \textquotedblleft image
charge\textquotedblright\ at $x=a_{1}+\Delta_{12}$. \ There is a final class
of $2-3$ string exchange diagrams which can also contribute to the alignment
of $\delta x_{1}$ when $N_{3}>0$. \ These correspond to diagrams with the same
shape as in the upper lefthand diagram of figure \ref{twoloop}, but with all
index loops distinct. \ In this case all three diagrams contribute with the
same sign to the alignment of $\delta x_{1}$ along the real axis:%
\begin{equation}
sign\left(  \frac{4}{W^{(2)}\left(  a_{1}\right)  \Delta_{12}\Delta_{13}%
}W^{(2)}(a_{1})\right)  \Rightarrow\delta x_{1}\in%
\mathbb{R}
\text{ \ \ \ \ \ for }c>a_{1}\text{.}%
\end{equation}
The above analysis indicates that the sign of each string exchange diagram
depends only on whether $c$ falls to the left or right of $x=a_{2}$ and
$x=a_{1}$. \ Indeed, the classical trajectories of the string exchange
processes in the complex $x$-plane sometimes must bend in order to avoid
passing too close to the $S^{2}$ at $x=c$.

\subsection{Alignment with Multiple Minimal $S^{2}$'s \label{ALIGNMULTIPLE}}

We now extend the analysis of the previous subsection to configurations where
$W^{\prime}(x)$ has multiple roots. \ Although it is tempting to speculate
based on the examples of the previous subsection that a large discrete jump in
the preferred alignment direction of a branch cut will not occur in highly
symmetric configurations, in this subsection we show that even when the roots
of $W^{\prime}(x)$ appear symmetrically along the real axis of the complex
$x$-plane, the orientation of the cuts in the lowest energy configuration
still suffers discrete jumps.

Most generally, we have:%
\begin{equation}
W^{\prime}(x)=g(x^{2}-a^{2})f(x)
\end{equation}
where $f(x)$ is a monic polynomial with isolated roots. \ With $N_{1}>0$
D5-branes located at $x=a>0$ and $N_{2}<0$ anti-D5-branes at $x=-a$ with a
vanishingly small number at the remaining roots of $W^{\prime}(x)$, the two
loop contribution to the energy density is:%
\begin{equation}
\delta E=\frac{\left\vert N_{1}\right\vert \left\vert N_{2}\right\vert }{4\pi
a^{2}}\operatorname{Re}\left(
\begin{array}
[c]{c}%
\left(  -3-2\underset{i=1}{\overset{m}{%
{\displaystyle\prod}
}}\frac{c_{i}-a}{a+c_{i}}-\underset{i=1}{\overset{m}{%
{\displaystyle\sum}
}}\frac{4a}{a-c_{i}}\right)  \left(  \delta x_{1}\right)  ^{2}\\
+\left(  -3-2\underset{i=1}{\overset{m}{%
{\displaystyle\prod}
}}\frac{a+c_{i}}{c_{i}-a}-\underset{i=1}{\overset{m}{%
{\displaystyle\sum}
}}\frac{4a}{a+c_{i}}\right)  \left(  \delta x_{2}\right)  ^{2}%
\end{array}
\right)
\end{equation}
where $c_{1},...,c_{m}$ denote the roots of $f(x)$. \ Treating the $c_{i}$ as
$m$ complex coordinates in $%
\mathbb{C}
^{m}$, note that the coefficients multiplying $\left(  \delta x_{1}\right)
^{2}$ and $\left(  \delta x_{2}\right)  ^{2}$ both vanish on complex manifolds
of dimension $m-1$. \ It follows that crossing this zero set will induce a
jump in the branch cut orientation of the energetically preferred confining vacuum.

Although an analysis indicating the various phase regions for cut alignment
for general $c_{i}$ will likely be complicated, when a higher degree of
symmetry is present, the analysis remains tractable. \ Restricting to $f(x)$
an even or odd polynomial so that:%
\begin{equation}
f(x)=x^{\sigma}\underset{i=1}{\overset{s}{%
{\displaystyle\prod}
}}(x^{2}-b_{i}^{2}) \label{ffunctiondefinition}%
\end{equation}
with $b_{i}$ all real and $\sigma=0$ or $1$, we find:%
\begin{equation}
\delta E=\frac{\left\vert N_{1}\right\vert \left\vert N_{2}\right\vert }{4\pi
a^{2}}\operatorname{Re}\left(  \left(  -5-\underset{i=1}{\overset{s}{\sum}%
}\frac{8}{1-\left(  b_{i}/a\right)  ^{2}}\right)  \left(  \left(  \delta
x_{1}\right)  ^{2}+\left(  \delta x_{2}\right)  ^{2}\right)  \right)  \text{.}%
\end{equation}
The alignment of cuts is parallel or perpendicular to the real axis of the
complex $x$-plane in the energetically preferred confining vacuum when:%
\begin{align}
5+\underset{i=1}{\overset{s}{\sum}}\frac{8}{1-\left(  b_{i}/a\right)  ^{2}}
&  >0\text{ \ \ \ \ \ \ \ \ \ \ \ \ \ (parallel)}\label{par}\\
5+\underset{i=1}{\overset{s}{\sum}}\frac{8}{1-\left(  b_{i}/a\right)  ^{2}}
&  <0\text{ \ \ \ \ \ (perpendicular).} \label{perp}%
\end{align}
Note that a perpendicular alignment of the cuts requires the existence of a
root of $f(x)$ such that $\left\vert b_{i}\right\vert >\left\vert a\right\vert
$. \ This suggests that in addition to the Coulomb attraction due to
brane/anti-brane forces, the geometry itself exerts a \textquotedblleft
torque\textquotedblright\ on the cuts. \ It is tempting to speculate that one
possible source of torque may be the condensation of D3-branes wrapping
collapsing $S^{3}$'s which are not threaded by flux.

As a final example, we consider the alignment of two branch cuts where the
background geometry is determined by the relation:%
\begin{equation}
W^{\prime}(x)=g\sin\frac{\pi x}{L}\text{.}%
\end{equation}
This corresponds to a geometry with an infinite one-dimensional lattice of
minimal size $S^{2}$'s located at $x=jL$ for all $j\in%
\mathbb{Z}
$. \ For configurations with $N_{1}$ D5-branes at $x=mL$ and $N_{2}$
anti-D5-branes at $x=0$, the two loop contribution to the vacuum energy
density is now:%
\begin{equation}
\delta E=\frac{\left\vert N_{1}\right\vert \left\vert N_{2}\right\vert }{\pi
m^{2}L^{2}}\operatorname{Re}\left(  \left(  -1+2(-1)^{m}\right)  \left(
\left(  \delta x_{1}\right)  ^{2}+\left(  \delta x_{2}\right)  ^{2}\right)
\right)  \text{.}%
\end{equation}
Thus, for $m$ even (resp. odd) the cuts align perpendicular (resp. parallel)
to the real axis of the complex $x$-plane.

\section{Geometry and Strong CP \label{StrongCP}}

Upon embedding our configuration in a compact Calabi-Yau, the effective
$\theta$-angle of the brane/anti-brane system becomes a dynamical field with
potential determined by the vacuum energy density of the open string system.
\ In this section we assume that the time-scale for the fluctuations of the
corresponding axion is sufficiently long compared to the time-scale associated
with the decay of a glueball phase to the energetically preferred confining
vacuum. \ The axion potential is then given by minimizing the vacuum energy
density over all of the confining vacua of the theory. \ The roots of
$W^{\prime}(x)$ determine the level of strong CP\ violation in each strongly
coupled gauge group. \ At one loop order the effective $\theta$-angle of the
$i^{\text{th}}$ D5-brane is defined by the relation:%
\begin{equation}
S_{i}=\Lambda_{i}^{3}=\zeta_{i}\left\vert \Lambda_{i}\right\vert ^{3}%
\exp\left(  i\theta_{i}/\left\vert N_{i}\right\vert \right)  \label{holIR}%
\end{equation}
where $\zeta_{i}$ denotes an $\left\vert N_{i}\right\vert ^{\mathrm{th}}$ root
of unity. \ The effective $\theta$-angles now follow from equations
(\ref{BCRIT}) and (\ref{BBCRIT}):%
\begin{align}
\theta_{i}  &  =\left\vert N_{i}\right\vert \arg W^{(2)}(a_{i})-\underset
{i\neq j\in B}{\sum}\left\vert N_{j}\right\vert \arg\Delta_{ij}^{2}%
+\underset{k\in\overline{B}}{\sum}\left\vert N_{k}\right\vert \arg\Delta
_{ik}^{2}+A\text{ \ \ \ \ (}N_{i}>0\text{)}\label{pos}\\
\theta_{i}  &  =\left\vert N_{i}\right\vert \arg W^{(2)}(a_{i})-\underset
{i\neq k\in\overline{B}}{\sum}\left\vert N_{j}\right\vert \arg\Delta_{ik}%
^{2}+\underset{j\in B}{\sum}\left\vert N_{k}\right\vert \arg\Delta_{ij}%
^{2}-A\text{\ \ \ \ \ (}N_{i}<0\text{)} \label{neg}%
\end{align}
where in the above we have absorbed all $\Lambda_{0}$ dependence into the
single variable $A$:%
\begin{equation}
A\equiv\underset{i}{\sum}N_{i}\arg\Lambda_{0}^{2}+\theta_{\mathrm{YM}}%
(\Lambda_{0})\text{.} \label{axion}%
\end{equation}
Note that upon rephasing the parameters $W^{(2)}(a_{i})$ and $\Delta_{ij}^{2}%
$, the value of $\theta_{i}$ also shifts. \ This is consistent with the
presence of the axial anomaly. \ The discrete symmetry CP is preserved by the
$i^{\text{th}}$ strongly coupled gauge group when $\theta_{i}$ vanishes.

Minimizing the vacuum energy density over all of the confining vacua yields
the axion potential for the effective $\theta$-angles:%
\begin{equation}
V_{\mathrm{ax}}=E^{(0)}+\min\left(  \delta E\right)
\end{equation}
where the $\min$ denotes minimization over all of the discrete glueball phases
and as before, $\delta E$ denotes the second term of equation
(\ref{twoloopshift}). \ From the perspective of the closed string dual, it is
natural to view $V_{\mathrm{ax}}$ as a function of $n$ independent variables
$\theta_{1},...,\theta_{n}$. \ On the other hand, from the perspective of the
open string description, there is a single $S^{2}$. \ It is therefore also
natural to treat the parameter $A$ defined in equation (\ref{axion}) as the
only dynamical axion of the theory. \ In subsections \ref{CPinvariant} and
\ref{discretesymmstrongcp} we determine constraints on the minima of
$V_{\mathrm{ax}}$ by further studying the behavior of $\delta E$ as a function
of the $\theta_{i}$'s as well as the $a_{i}$'s.

\subsection{CP\ Invariant Submanifolds\label{CPinvariant}}

Viewing the locations of the minimal size $S^{2}$'s as an $n$-component vector
$\overrightarrow{a}=(a_{1},...,a_{n})\in%
\mathbb{C}
^{n}$, we now argue that when $l$ of the $S^{2}$'s are occupied by a
collection of branes and anti-branes and $n-l$ is sufficiently large, there
exists a real submanifold of points in $%
\mathbb{C}
^{n}$ with real codimension $l$ such that the corresponding axion potential
achieves a minimum at $\theta_{i}=0$ for all $i$. \ It follows from equation
(\ref{twoloopshift}) and the explicit expressions of appendix A that the two
loop contribution to the vacuum energy density from terms linear in $S_{i}$ is
schematically of the form:%
\begin{equation}
\delta E=\underset{i\text{ occupied}}{\sum}\operatorname{Re}\left[
C_{i}(\overrightarrow{a},\overrightarrow{N})S_{i}\right]
\end{equation}
where the $C_{i}(\overrightarrow{a},\overrightarrow{N})$ are rational
functions of the $a_{i}$. \ $\delta E$ has a minimum at $\theta_{i}=0$ when
$C_{i}<0$. \ The requirement that all branes and anti-branes preserve strong
CP imposes $l$ real conditions on the parameter space of the $a_{i}$'s which
will generically be satisfied provided $n-l$ is sufficiently large. \ Hence,
there is a manifold of points of real dimension $2n-l$ contained in $%
\mathbb{C}
^{n}$ such that all branes preserve strong CP. \ Note in particular that this
result is independent of whether we treat all of the $\theta_{i}$'s or simply
$A$ as dynamical fields.

\subsection{Discrete Symmetries and Strong CP \label{discretesymmstrongcp}}

We now show that discrete symmetries of the Calabi-Yau translate into
constraints on the amount of strong\ CP\ violation. \ Letting $\sigma$ denote
the action of the permutation symmetry on the labelling of the branes, we
consider brane configurations with $W^{\prime}(x)$ a polynomial with real
coefficients such that $a_{i}=\overline{a_{j}}\equiv a_{\sigma(i)}$ implies
$N_{i}=N_{j}$. \ Returning to equation (\ref{twoloopshift}) and the explicit
expression for $\delta\tau$ given in appendix A, the coefficients multiplying
$S_{i}$ are complex conjugates of the coefficients in the analogous
contribution from $S_{\sigma(i)}$. \ To establish this, we consider without
loss of generality the $\theta$-angle for $N_{i}>0$ D5-branes at $x=a_{i}$.
\ Note that:%
\begin{align}
Q_{i}  &  \equiv\underset{j\in\overline{B}}{%
{\displaystyle\sum}
}\left\vert N_{i}\right\vert \left\vert N_{j}\right\vert \frac{1}%
{W^{(2)}\left(  a_{i}\right)  }\frac{2}{\Delta_{ij}^{2}}\left(  -1-\Delta
_{ij}\frac{W^{(3)}\left(  a_{i}\right)  }{W^{(2)}\left(  a_{i}\right)
}+2\frac{W^{(2)}(a_{i})}{W^{(2)}(a_{j})}\right)  =\overline{Q_{\sigma(i)}}\\
P_{i}  &  \equiv\underset{i\neq s\in B,t\in\overline{B}}{%
{\displaystyle\sum}
}\left\vert N_{s}\right\vert \left\vert N_{t}\right\vert \left(  \frac
{4}{W^{(2)}(a_{s})}\frac{1}{\Delta_{si}}\frac{1}{\Delta_{st}}%
+(t\leftrightarrow s)+(i\leftrightarrow s)\right)  =\overline{P_{\sigma(i)}%
}\text{.}%
\end{align}
The net contribution from terms linear in $S_{i}$ and $S_{\sigma(i)}$ is
therefore:%
\begin{align}
\delta E\left(  \theta_{i}\right)   &  =\operatorname{Re}\left(  C_{i}%
S_{i}\right) \label{top}\\
\delta E\left(  \theta_{\sigma(i)}\right)   &  =\operatorname{Re}\left(
\overline{C_{i}}S_{\sigma(i)}\right)  \label{bot}%
\end{align}
with $C_{i}$ a complex number.

We now study the minima of the axion potential first in the case where the
$\theta_{i}$'s are all dynamical fields, and then in the case where only $A$
is a dynamical field. \ It follows from equations (\ref{top}) and (\ref{bot})
that in the first case the minimum energy configuration satisfies
$\left\langle \theta_{i}\right\rangle =-\left\langle \theta_{\sigma
(i)}\right\rangle $. \ In particular, when $a_{i}$ is purely real, the
effective $\theta$-angle vanishes. \ Next treat $V_{\mathrm{ax}}$ as a
function of the single field $A$. \ In this case, the symmetries of the system
in tandem with equation (\ref{pos}) yield the constraint:%
\begin{equation}
\theta_{i}-A=A-\theta_{\sigma(i)}%
\end{equation}
so that $A=\theta_{i}$ when $a_{i}$ is purely real. \ The combined
contribution to the energy density from $\theta_{i}$ and $\theta_{\sigma(i)}$
now follows from equations (\ref{top}) and (\ref{bot}):%
\begin{equation}
\delta E\left(  \theta_{i}\right)  +\delta E\left(  \theta_{\sigma(i)}\right)
=\left\vert C_{i}\right\vert \left\vert S_{i}\right\vert \left(  \cos\left(
\gamma_{i}+\frac{A+2\pi k_{i}}{\left\vert N_{i}\right\vert }\right)
+\cos\left(  -\gamma_{i}+\frac{A+2\pi k_{\sigma(i)}}{\left\vert N_{i}%
\right\vert }\right)  \right)
\end{equation}
where in the above,%
\begin{equation}
\gamma_{i}=\arg C_{i}+\arg S_{i}-\frac{A}{\left\vert N_{i}\right\vert
}\text{.}%
\end{equation}
Minimizing with respect to $k_{i}$ and $k_{\sigma(i)}$, the corresponding
axion potential has a minimum at $A=0$.

\section{Critical Points and Unoccupied $S^{2}$'s\label{Z2symm}}

To further study the phase structure of brane/anti-brane configurations
wrapping $S^{2}$'s in the presence of additional minimal size $S^{2}$'s, we
first determine the behavior of the critical points of the corresponding
effective potential. \ As shown in \cite{HSV}, although it is difficult to
directly solve for the critical points of the $n$-cut effective potential
$V_{\mathrm{eff}}$, it is mathematically simpler to solve for the fluxes as a
function of the critical points of $V_{\mathrm{eff}}$. \ In \cite{HSV} these
\textquotedblleft attractor-like equations\textquotedblright\ were used to
establish the gross features of multi-cut geometries. \ It follows from
section \ref{Unoccupied} that the two cut attractor-like equations extend to
geometries where only two minimal size $S^{2}$'s are wrapped by branes and
anti-branes. \ We now review the two cut attractor-like equations and then
apply them in a special class of geometries and flux configurations where the
analysis of the critical points remains tractable. \ These results are then
used in section \ref{Breakdown} to show that the breakdown in metastability is
similar to the two cut geometries studied in \cite{HSV}.

The critical points of the two cut system satisfy two two-component vector
relations \cite{HSV}:%
\begin{align}
\frac{2i}{C}\left[
\begin{array}
[c]{c}%
N_{1}\\
N_{2}%
\end{array}
\right]   &  =\left[
\begin{array}
[c]{cc}%
\rho_{w} & 1\\
\overline{\rho_{v}}\rho_{w} & \overline{\rho_{v}}%
\end{array}
\right]  \tau\left[
\begin{array}
[c]{c}%
1\\
-1
\end{array}
\right]  -\left[
\begin{array}
[c]{cc}%
\rho_{w} & \overline{\rho_{v}}\rho_{w}\\
1 & \overline{\rho_{v}}%
\end{array}
\right]  \overline{\tau}\left[
\begin{array}
[c]{c}%
1\\
-1
\end{array}
\right] \label{attractorone}\\
\frac{2i}{C}\left[
\begin{array}
[c]{c}%
\alpha\\
\alpha
\end{array}
\right]   &  =\tau\left[
\begin{array}
[c]{cc}%
\rho_{w} & \overline{\rho_{v}}\rho_{w}\\
1 & \overline{\rho_{v}}%
\end{array}
\right]  \overline{\tau}\left[
\begin{array}
[c]{c}%
1\\
-1
\end{array}
\right]  -\overline{\tau}\left[
\begin{array}
[c]{cc}%
\rho_{w} & 1\\
\overline{\rho_{v}}\rho_{w} & \overline{\rho_{v}}%
\end{array}
\right]  \tau\left[
\begin{array}
[c]{c}%
1\\
-1
\end{array}
\right]  \label{attractor2}%
\end{align}
where $\tau$ is the $2\times2$ period matrix, $C$ is a constant non-zero
complex number, and we have introduced:%
\begin{align}
\rho_{v}  &  =-\frac{d_{3}\pm\sqrt{d_{3}^{2}-4d_{1}d_{2}}}{2d_{2}%
}\label{nuequation}\\
\rho_{w}  &  =-\frac{d_{3}\pm\sqrt{d_{3}^{2}-4d_{1}d_{2}}}{2d_{1}}
\label{omegaequation}%
\end{align}
with:%
\begin{equation}
d_{1}=\det\left[
\begin{array}
[c]{cc}%
\mathcal{F}_{111} & \mathcal{F}_{112}\\
\mathcal{F}_{112} & \mathcal{F}_{122}%
\end{array}
\right]  ,\text{ }d_{2}=\det\left[
\begin{array}
[c]{cc}%
\mathcal{F}_{112} & \mathcal{F}_{122}\\
\mathcal{F}_{122} & \mathcal{F}_{222}%
\end{array}
\right]  ,\text{ }d_{3}=\det\left[
\begin{array}
[c]{cc}%
\mathcal{F}_{111} & \mathcal{F}_{112}\\
\mathcal{F}_{122} & \mathcal{F}_{222}%
\end{array}
\right]  . \label{dets}%
\end{equation}
The $\pm$ signs of equations (\ref{nuequation}) and (\ref{omegaequation}) are
correlated. \ The requirement that $g_{YM}^{2}>0$ leads to an unambiguous
assignment of brane type for each branch. \ Switching from the $+$ to the $-$
branch of equations (\ref{nuequation}) and (\ref{omegaequation}) changes all
branes (anti-branes) into anti-branes (branes).

Next consider more general geometries with:%
\begin{equation}
W^{\prime}(x)=g(x^{2}-a^{2})f(x) \label{geom}%
\end{equation}
where $g,a>0$ and $f(x)$ is an even or odd polynomial whose roots are all
isolated and real. \ This corresponds to a geometry where the minimal size
$S^{2}$'s are situated along the real axis of the complex $x$-plane and are
symmetrically arranged with respect to the imaginary axis. \ We take $N_{1}>0$
branes at $x=a_{1}=a$ and $N_{2}<0$ anti-branes at $x=a_{2}=-a$ with the
remaining minimal size $S^{2}$'s left unwrapped by branes.

We now show that when $N_{1}=-N_{2}$, the corresponding effective potential
admits critical points such that $t_{1}=\overline{t_{2}}$. \ The analysis is
very similar to that given in \cite{HSV} and we will therefore only summarize
the various steps in the computation. \ It follows from the $%
\mathbb{Z}
_{2}$ symmetry of the configuration that there exists a finite neighborhood
around $t_{1},t_{2}=0$ such that for $t_{1}=\overline{t_{2}}$ we have:%
\begin{align}
\left[
\begin{array}
[c]{cc}%
\overline{\tau_{11}} & \overline{\tau_{12}}\\
\overline{\tau_{12}} & \overline{\tau_{22}}%
\end{array}
\right]   &  =\left[
\begin{array}
[c]{cc}%
-\tau_{22} & -\tau_{12}\\
-\tau_{12} & -\tau_{11}%
\end{array}
\right]  -\frac{M-\overline{M}}{2\pi i}\left[
\begin{array}
[c]{cc}%
1 & 1\\
1 & 1
\end{array}
\right] \\
\mathcal{F}_{111}  &  =(-1)^{\deg f}\overline{\mathcal{F}_{222}}\\
\mathcal{F}_{112}  &  =(-1)^{\deg f}\overline{\mathcal{F}_{122}}%
\end{align}
where $M\equiv\log\left(  \Lambda_{0}^{2}/\Delta^{2}\right)  $. \ This implies
$d_{1}=\overline{d_{2}}$, $d_{3}=\overline{d_{3}}$ and thus, $\overline
{\rho_{v}}=\rho_{w}$. \ Returning to equations (\ref{attractorone}) and
(\ref{attractor2}), $t_{1}=\overline{t_{2}}$ implies $N_{1}=-N_{2}$ and:%
\begin{equation}
\frac{\alpha}{N_{1}}=\frac{1+\rho_{w}}{1-\rho_{w}}\left(  \tau_{11}+\tau
_{12}+\frac{M-\overline{M}}{2\pi i}\right)  +\frac{\tau_{22}-\tau_{11}}%
{1-\rho_{w}}\text{.} \label{anrat}%
\end{equation}

\section{Breakdown of Metastability Revisited \label{Breakdown}}

In two cut geometries with $N_{1}=-N_{2}$ and both branch cuts aligned along
the real axis of the complex $x$-plane, there is a critical amount of flux
beyond which the effective potential ceases to admit critical points for
$t_{i}$ small \cite{HSV}. \ Viewing the ratio $\alpha/N_{1}$ as a function of
the real modulus $t\equiv t_{1}=t_{2}>0$, a critical point of this function
corresponds to the merging of a local minimum and maximum of $V_{\mathrm{eff}%
}$. \ In this section we study the analogous situation when additional minimal
size $S^{2}$'s are present in the open string description of the system.

Based on the condition for a breakdown in metastability in two cut geometries
given by equation (\ref{breakdown}), our expectation is that two loop effects
will destabilize the metastable vacua when the one loop vacuum energy density
becomes comparable to the energy density of a supersymmetric brane
configuration. \ Testing this expectation requires determining the
attractor-like equations for more general flux configurations and then
determining when the effective potential experiences a merger of two critical
points. \ Although for general configurations this appears to be intractable
with present techniques, when the branes and anti-branes of the open string
description cluster into two groups so that the system is characterized by the
brane/anti-brane distance $\Delta_{\mathrm{sep}}$ and the internal separation
scale within a cluster $\Delta_{\mathrm{inter}}$ with $\left\vert
\Delta_{\mathrm{inter}}\right\vert \ll\left\vert \Delta_{\mathrm{sep}%
}\right\vert $, the analysis reduces to that of the two cut system. \ Indeed,
it follows from general renormalization group arguments that at energy scales
above the mass scale set by $\Delta_{\mathrm{inter}}$, each cluster of branes
will behave as a single stack of \textquotedblleft unHiggsed\textquotedblright%
\ branes. \ In this case the expected breakdown in metastability is given by
the general bound for the two cut system:%
\begin{equation}
\frac{8\pi}{g_{YM}^{2}}\left(  \left\vert N_{1}\right\vert +\left\vert
N_{2}\right\vert \right)  -\frac{2}{\pi}\left\vert N_{1}\right\vert \left\vert
N_{2}\right\vert \log\left\vert \frac{\Lambda_{0}}{\Delta_{\mathrm{sep}}%
}\right\vert ^{2}\gtrsim\frac{8\pi}{g_{YM}^{2}}\left\vert N_{1}+N_{2}%
\right\vert
\end{equation}
where $N_{1}>0$ denotes the net number of branes from the first cluster and
$N_{2}<0$ denotes the net number of anti-branes from the second cluster.
\ This is indeed consistent with expectations based on treating the branes in
the geometry as a distribution of point charges. \ For more general
configurations the precise bound on the one loop contribution is more involved
because the channel of brane/anti-brane annihilation depends on the relative
positions of all branes and anti-branes in the geometry. \ This complicates
the estimate of when to expect a breakdown in metastability.

To avoid such subtleties, in the rest of this section we restrict our analysis
to configurations with $N_{1}=N>0$ D5-branes at $x=a_{1}=a>0$ and $N_{2}=-N$
anti-D5-branes at $x=a_{2}=-a$ in geometries with:%
\begin{equation}
W^{\prime}(x)=g(x^{2}-a^{2})f(x)
\end{equation}
and $f(x)$ as in equation (\ref{ffunctiondefinition}). \ It follows from the
results of section \ref{Z2symm} that in this case there exist critical points
which satisfy $t_{1}=\overline{t_{2}}\equiv t$. \ Using the explicit
expressions for $\tau_{ij}$ and $\mathcal{F}_{ijk}$ given in appendix B,
expanding equation (\ref{anrat}) to lowest order in $r>0$ with $t=e^{i\phi}r$
yields:%
\begin{equation}
\frac{2\pi i\alpha}{N}=\pm\left(  1+2(-1)^{\deg f}e^{\pm i\phi}\beta r\right)
\left(  \log\left(  re^{i\phi}\right)  -\log\left\vert \frac{\Lambda_{0}%
}{\Delta}\right\vert ^{4}\right)  +\mathcal{O}(r)+\mathcal{O}(r^{2}\log r)
\label{expandedanrat}%
\end{equation}
where the variable $\beta$ is defined by equation (\ref{betadef}) in appendix
B. \ Because the results of subsection \ref{ALIGNMULTIPLE} establish that the
minimal energy confining vacuum corresponds to a configuration where the
branch cuts align parallel or perpendicular to the real axis of the complex
$x$-plane, it is enough to study the critical points of $2\pi i\alpha/N$ as a
function of $t\in%
\mathbb{R}
$. \ Further, due to the fact that each value of $\theta_{YM}$ will be
attained by some value of $r$ by changing $\phi$ by a small amount, it is
enough to demonstrate that $\operatorname{Re}[2\pi i\alpha/N]$ has an extremum
as a function of $r$ for $\phi=0$ or $\pi$.

At leading order in $r$, the critical points of $\operatorname{Re}\left[  2\pi
i\alpha/N\right]  $ satisfy:%
\begin{equation}
\frac{1}{r_{\ast}}=-2(-1)^{\vartheta+\deg f}\beta\log\left(  r_{\ast
}\left\vert \frac{\Delta}{\Lambda_{0}}\right\vert ^{4}\right)
\label{critequation}%
\end{equation}
where the variable $\vartheta=0$ for $\phi=0$ and $\vartheta=1$ for $\phi=\pi
$. \ The existence of a critical point therefore requires $(-1)^{\vartheta
+\deg f}\beta>0$. \ Returning to the definition of $\beta$ given by equation
(\ref{betadef}) in appendix B, note that this inequality is in accord with the
condition for parallel ($\vartheta=0$) and perpendicular $(\vartheta=1$)
alignment given by equations (\ref{par}) and (\ref{perp}), respectively.

The critical value of the 't Hooft coupling $\lambda=g_{YM}^{2}N$ for which
the system is no longer metastable is given by evaluating equation
(\ref{expandedanrat}) at the value determined by equation (\ref{critequation}%
). \ Dropping all subleading logarithms, a crude estimate of this value is:%
\begin{equation}
\frac{8\pi^{2}}{\lambda_{\ast}}=\log\left\vert \frac{\Lambda_{0}}{\Delta
}\right\vert ^{4}+\log\left\vert 20+\underset{i=1}{\overset{s}{\sum}}\frac
{32}{1-\left(  b_{i}/a\right)  ^{2}}\right\vert \text{.}%
\end{equation}

\subsection{Endpoints and Further Transitions \label{nonkahler}}

As shown in \cite{HSV} for the two cut system, once metastability is lost, the
sizes of the cuts will expand in order to facilitate the annihilation of flux
lines. \ This expansion leads to the collapse of a 3-cycle in the geometry, at
which point additional light degrees of freedom enter the low energy spectrum.
\ Whereas the presence of D5-branes wrapping this collapsed 3-cycle tend to
relax the system back to a metastable geometry of lower flux, the presence of
a D3-brane wrapping this 3-cycle can drive the system to a non-K\"{a}hler geometry.

Even though the alignment of cuts for metastable vacua is sometimes
perpendicular and sometimes parallel to the real axis of the complex
$x$-plane, in all cases that we have studied numerically the mode of
instability drives the cuts to align and expand along the real axis of the
complex $x$-plane once metastability is lost.

As opposed to the two cut geometry, however, for more general flux
configurations a branch cut supported by flux may first collide with another
cut supported by flux of the same sign. \ In this case, the presence of
D5-branes wrapping the collapsing 3-cycle separate vacua in the four
dimensional spacetime which are nearly degenerate in energy density. \ Hence,
in comparison to the two cut geometry, a D3-brane wrapping this same collapsed
cycle will play a more pronounced r\^{o}le in determining the endpoint of this
transition. \ Note, however, that generic values of $\theta_{\mathrm{YM}}$ can
also prevent such collisions. \ Indeed, whereas $\theta_{\mathrm{YM}}$ tends
to tip branch cuts supported by positive and negative RR fluxes so that they
will continue to collide, when both branch cuts support flux of the same sign,
they tip so as to remain parallel.

\section{Modes of Annihilation\label{DecayModes}}

When $W^{\prime}(x)$ has two roots, the branes and anti-branes eventually
annihilate by traversing an interpolating minimal size 3-chain. \ In this
section we demonstrate that for more general geometries, new modes of
brane/anti-brane annihilation appear or disappear as the locations of the
minimal size $S^{2}$'s change.

This is mathematically similar to the problem of determining the BPS spectrum
of $\mathcal{N}=2$ gauge theories geometrically engineered in type IIB\ string
theory on geometries of the form:%
\begin{equation}
y^{2}=P(x)+uv
\end{equation}
where $P(x)$ is a polynomial with isolated roots. \ The spectrum of solitons
given by D3-branes wrapping supersymmetric 3-cycles which interpolate between
the roots of $P(x)$ depends strongly on the relative positions of these roots
\cite{ShapereVafa}. \ As a particular example, when the roots of $P(x)$ are
all located on the real axis of the complex $x$-plane and ordered as
$r_{n}<...<r_{1}$, the BPS soliton\ connecting $r_{n}$ to $r_{1}$ decays into
a collection of $n-1$ BPS\ solitons joining $r_{i}$ to $r_{i-1}$ for $1<i<n+1$
\cite{ShapereVafa}.

We are interested in a special limit of the above analysis where $P(x)\simeq
W^{\prime}(x)^{2}$. \ In this case the BPS\ spectrum will be similar to that
determined by two dimensional $\mathcal{N}=(2,2)$ Landau-Ginzburg theory
\cite{ShapereVafa}. \ It follows from arguments similar to those presented in
\cite{ShapereVafa} that the spectrum of nearly BPS D5-branes wrapping 3-cycles
will develop instabilities as the roots of $W^{\prime}(x)$ change.

In the next two subsections we restrict our analysis for illustrative purposes
to geometries with three minimal size $S^{2}$'s. \ We determine the wall of
marginal stability for a D5-brane domain wall solution wrapping a 3-chain
connecting two roots of $W^{\prime}(x)$ to decay to a two domain wall process.
\ We next demonstrate that such solutions have minimal tension. \ This implies
that for more general geometries, semi-classical brane/anti-brane annihilation
proceeds via a multi-stage \textquotedblleft hopping\textquotedblright%
\ process. \ Finally, in subsection \ref{HOPPING} we indicate some qualitative
features of this phenomenon.

\subsection{Spectrum of Nearly BPS\ Domain Walls with Three Minimal $S^{2}$'s}

We now determine the nearly BPS\ spectrum of D5-branes wrapping interpolating
$3$-chains in geometries with three minimal size $S^{2}$'s. \ Given a
$3$-chain interpolating between $x=a_{i}$ and $x=a_{j}$, the tension of the
corresponding domain wall is well-approximated by:%
\begin{equation}
T\left[  \gamma_{ij}\right]  =\frac{1}{g_{s}}\underset{\gamma_{ij}}{\int
}\left\vert W^{\prime}(x)\right\vert \left\vert dx\right\vert
\label{truetension}%
\end{equation}
where $\gamma_{ij}$ denotes a contour in the complex $x$-plane. \ For a single
nearly BPS domain wall, the tension is:%
\begin{equation}
T_{ij}=\frac{1}{g_{s}}\left\vert W\left(  a_{i}\right)  -W\left(
a_{j}\right)  \right\vert \label{BPStension}%
\end{equation}
corresponding to a path in the complex $x$-plane such that the phase of
$W^{\prime}(x)$ remains constant for all $x\in\gamma_{ij}$.

\begin{figure}
[ptb]
\begin{center}
\includegraphics[
height=2.0003in,
width=3in
]%
{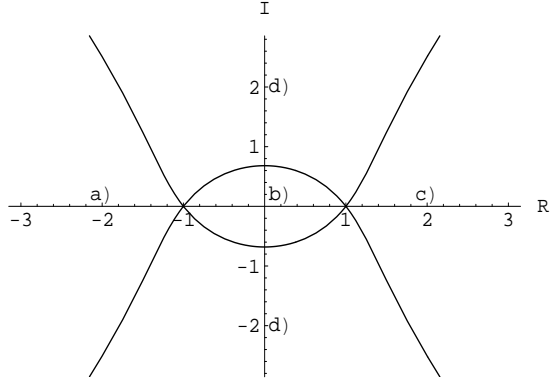}%
\caption{Plot of the walls of marginal stability with three minimal size
$S^{2}$'s located at $a_{1}=1$, $a_{2}=-1$ and $a_{3}=R+iI$. \ As the point
$a_{3}$ is varied throughout the $R-I$ plane, a single BPS\ domain wall
solution connecting two of the $a_{i}$'s may decay into a two stage process.
\ When $a_{3}$ is located in region a), the process $a_{3}\rightarrow a_{1}$
decays. \ For $a_{3}$ in region b), the process $a_{2}\rightarrow a_{1}$
decays and in region c) the process $a_{2}\rightarrow a_{3}$ decays. \ In
region d) all processes of the form $a_{i}\rightarrow a_{j}$ are allowed.}%
\label{mswall}%
\end{center}
\end{figure}

An important caveat to equation (\ref{BPStension}) is that it assumes the
existence of a single nearly BPS\ domain wall. \ Letting $a_{1}=-a_{2}=a>0$
and $a_{3}=c$, note that $c=0$ implies $W(a)=W(-a)$ so that the purported
domain wall solution connecting $a$ and $-a$ would have zero tension. \ This
unphysical result establishes the absence of such a single BPS domain wall in
this region of configuration space. \ We now determine the wall of marginal
stability for a single BPS\ domain wall to decay into two constituent
products. \ The central charge of each candidate BPS domain wall is:%
\begin{equation}
Z_{ij}=W\left(  a_{j}\right)  -W\left(  a_{i}\right)  \text{.}%
\end{equation}
The wall of marginal stability corresponds to the locus of non-trivial $c$
values such that the phases of the central charges for the candidate decay
products align. \ Setting $c/a\equiv R+iI$, the defining equation for the wall
of marginal stability is:%
\begin{equation}
I_{MS}(R)=\pm\sqrt{R^{2}-3+2\sqrt{R^{4}-3R^{2}+3}}\text{.} \label{MSCondition}%
\end{equation}
See figure \ref{mswall} for the region of marginal stability in the $R$-$I$
plane. Assuming that the brane/anti-brane configuration spontaneously breaks
$\mathcal{N}=2$ supersymmetry, the overall shape of this wall of marginal
stability should remain correct up to $1/N$ corrections.%

\subsection{Tension Minimizing Solutions}

When a single nearly BPS\ domain wall solution ceases to exist, the total
tension of the resulting decay products is a global minimum of the tension
formula given by equation (\ref{truetension}). \ To establish this, let $x(t)$
denote a parametrization of the path $\gamma_{12}$ which runs from
$x(0)=a_{1}$ to $x(1)=a_{2}$. \ It follows from the Cauchy-Riemann equations
that the net tension of the D5-brane wrapping the corresponding 3-chain is:%
\begin{equation}
T\left[  \gamma_{12}\right]  =\frac{1}{g_{s}}\underset{0}{\overset{1}{\int}%
}\left\vert \left(  \dot{W}_{1}\right)  ^{2}+\left(  \dot{W}_{2}\right)
^{2}\right\vert ^{1/2}dt \label{TENSION}%
\end{equation}
where we have decomposed $W$ into real and imaginary parts so that
$W=W_{1}+iW_{2}$. \ Equation (\ref{TENSION}) implies that minimum tension
paths in the complex $x$-plane map to piecewise straight line paths in the
complex $W$-plane.

To demonstrate that the two product process described in the previous
subsection minimizes $T$, we first restrict to the case with $a_{3}=iL$ and
$L\in%
\mathbb{R}
$. \ In the complex $W$-plane the decay products of the domain wall solution
connecting $a$ to $-a$ map to a straight line connecting $W(a)$ to $W(iL)$ and
a further straight line connecting $W(iL)$ to $W(-a)$. \ The values of $W(\pm
a)$ and $W(iL)$ in the complex $W$-plane are:%
\begin{align}
W\left(  \pm a\right)   &  =\frac{g}{12}a^{4}\left(  -3\pm8i\frac{L}{a}\right)
\\
W(iL)  &  =\frac{g}{12}L^{4}\left(  -1-6\frac{a^{2}}{L^{2}}\right)
\end{align}
so that for $g>0$, the locations of $W\left(  \pm a\right)  $ correspond to
reflections across the real axis of the complex $W$-plane. \ On the other
hand, for $-\left\vert I_{MS}\left(  R=0\right)  \right\vert <L<\left\vert
I_{MS}\left(  R=0\right)  \right\vert $, $\operatorname{Re}\left[
W(iL)\right]  >\operatorname{Re}\left[  W(\pm a)\right]  $ so that $W(iL)$
lies to the right of $W(\pm a)$ in the complex $W$-plane. \ While the shortest
path between the points $W(a)$ and $W(-a)$ in the complex $W$-plane
corresponds to a straight vertical line, this may not correspond to a path in
the complex $x$-plane. \ Indeed, consider any path in the complex $x$-plane
connecting $x=a$ to $x=-a$. \ Such a path will necessarily attain the value
$x=iy$ for some $y\in%
\mathbb{R}
$. \ In the complex $W$-plane, the path must therefore pass through the value
$W(iy)$. \ Comparing the values of $W(iy)$ and $W(iL)$, we find:%
\begin{equation}
W\left(  iy\right)  -W(iL)=\frac{g}{12}\left(  L-y\right)  ^{2}\left(
6a^{2}+2y^{2}+(L+y)^{2}\right)  >0\text{.}%
\end{equation}
Hence, when $\operatorname{Re}\left[  W(iL)\right]  >\operatorname{Re}\left[
W(\pm a)\right]  $, the piecewise straight line path in the complex $W$-plane
will be shortest when $y=L$. \ Performing an arbitrary real deformation of
$a_{3}$ away from the imaginary axis, it now follows that for all points which
lie within the region bounded by the marginal stability curves given by
equation (\ref{MSCondition}), the decay products of the single nearly BPS
domain wall solution minimize the tension functional of equation
(\ref{truetension}).

\subsection{Hopping Effects \label{HOPPING}}

\begin{figure}
[ptb]
\begin{center}
\includegraphics[
height=3.1903in,
width=5.6853in
]%
{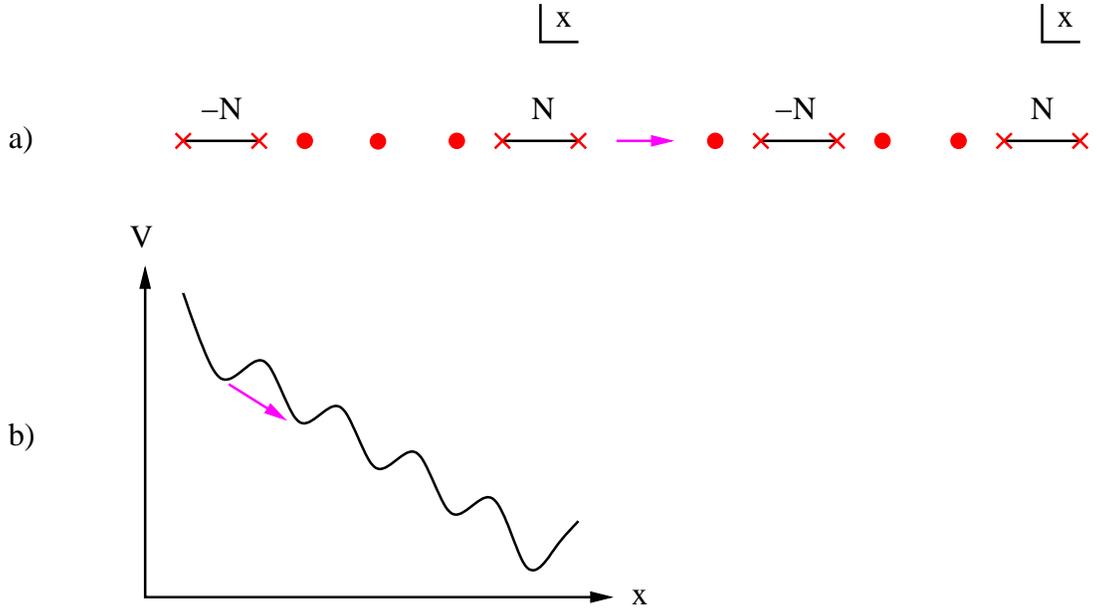}%
\caption{The positions of the roots of $W^{\prime}(x)$ can sometimes obstruct
flux line annihilation in the closed string dual geometry. \ Instead, the flux
must perform discrete jumps to other cuts before annihilating. \ In (a) a
hopping event where all $-N$ units of flux transfer to another cut is shown.
\ The effective potential as a function of $x\in\mathbb{R} $ seen by the
branes is shown in (b).}%
\label{cutwave}%
\end{center}
\end{figure}

Upon wrapping branes and anti-branes on two minimal size $S^{2}$'s, the
results of the previous subsection show that the additional $S^{2}$'s present
in the geometry can sometimes obstruct direct brane/anti-brane annihilation.
\ In a semi-classical $n$-domain wall decay process, a collection of branes
tunnels from one minimal size $S^{2}$ to the next until finally annihilating
with an anti-brane at the $n^{\text{th}}$ stage. \ In the closed string dual
description, flux lines hop from cut to cut before annihilating. \ The system
lowers its energy during each intermediate stage due to the one loop
contribution to the vacuum energy density from terms proportional to
$\log\left\vert a_{\mathrm{brane}}-a_{\mathrm{anti-brane}}\right\vert $.
\ Because the tensions of the domain wall solutions are all comparable, the
rate of decay due to hopping is much slower than direct annihilation. \ See
figure \ref{cutwave} for a depiction of this phenomenon. \ In this subsection
we discuss some further consequences which follow from hopping effects.%

Letting $\Gamma_{i\rightarrow i+1}$ denote the decay rate per unit volume for
branes to tunnel from $x=a_{i}$ to $x=a_{i+1}$, the rate for an $n$-stage
annihilation event which begins with $i=1$ and ends with $i=n+1$ is given by
summing over the inverse rates of each individual link in the decay
chain\footnote{In this formula we assume that each successive bubble of vacuum
can nucleate anywhere within the four dimensional Minkowski spacetime. \ This
provides an adequate approximation of the decay of the false vacuum because
the decay rate of each term in the decay chain is exponentially suppressed.}:%
\begin{equation}
\Gamma_{\mathrm{net}}^{-1}=\underset{i=1}{\overset{n}{\sum}}\Gamma
_{i\rightarrow i+1}^{-1}%
\end{equation}
so that a single hopping event can significantly retard the time-scale for a
drop in the total number of branes in the system.

Recall that in the dual closed string description there is a critical amount
of flux which can thread an $S^{3}$ before metastability is lost. \ It follows
that in certain cases a hopping event can trigger a breakdown in metastability
when too much flux threads an $S^{3}$. \ Combining this with the discussion in
subsection \ref{nonkahler}, this provides a completely dynamical mechanism
whereby a system can begin in a metastable Calabi-Yau geometry and transition
to a non-K\"{a}hler geometry.

As a collection of branes hops from one site to the next, the energetically
preferred direction of alignment for the branch cuts can also change. \ As a
simple example, consider the alignment of branch cuts in the geometry given by
an infinite one-dimensional lattice of evenly spaced minimal size $S^{2}$'s of
the type studied in subsection \ref{ALIGNMULTIPLE}. \ For a pair of branes and
anti-branes separated by $m$ lattice sites, the associated branch cuts align
parallel to the real axis for $m$ odd, and perpendicular to the real axis for
$m$ even. \ As the branes begin to hop towards one another, the energetically
preferred orientation of the cuts will also change.

The minima of the axion potential determined by equation (\ref{twoloopshift})
will also change. \ Indeed, if the number of occupied minimal size $S^{2}$'s
changes from $l$ to $l^{\prime}$, it follows from the analysis of subsection
\ref{CPinvariant} that when $n-l$ and $n-l^{\prime}$ are both sufficiently
large, the manifold of points $(a_{1},...,a_{n})\in%
\mathbb{C}
^{n}$ which preserve strong CP will change from real dimension $2n-l$ to
$2n-l^{\prime}$.

\section{Radial Mode Stabilization and Glueball Phases\label{Radion}}

In this section we propose a speculative mechanism which may stabilize the
radial mode in \textit{compact} Calabi-Yau threefolds. \ Interpreting the two
loop contribution to the vacuum energy density as a radial potential, we now
show that when the phases of the glueball fields anti-align with the
energetically preferred confining vacuum, this potential contains a power law
repulsion term which can counter the Coulomb attraction term present at one loop.

For simplicity we first consider a two cut configuration with $N_{1}=N$
D5-branes and $N_{2}=-N$ anti-D5-branes separated by a distance $\Delta
=a_{1}-a_{2}$. \ Treating $\theta_{\mathrm{YM}}$ as a mode which has already
developed an expectation value, we identify the two loop corrected value of
the vacuum energy density with the radial potential:%
\begin{align}
V_{\mathrm{rad}}\left(  \Delta\right)   &  =\frac{16\pi N}{g_{\mathrm{YM}}%
^{2}}-\frac{2N^{2}}{\pi}\log\left\vert \frac{\Lambda_{0}}{\Delta}\right\vert
^{2}\\
&  -\frac{20N^{2}}{\pi}\left\vert \frac{\Lambda_{0}}{\Delta}\right\vert
^{4}e^{-8\pi^{2}/\lambda}\left(  \cos\frac{2\pi k+\theta_{\mathrm{YM}}}%
{N}+\cos\frac{2\pi l+\theta_{\mathrm{YM}}}{N}\right)
\end{align}
where we have introduced the 't Hooft coupling $\lambda=g_{\mathrm{YM}}^{2}N$
and $k$ and $l$ are integers which label the distinct confining vacua of the
system. \ Note that whereas the second term produces a logarithmic attraction
between the branes and anti-branes, the sign of the power law contribution to
$V_{\mathrm{rad}}\left(  \Delta\right)  $ crucially depends on the particular
confining vacuum of the system. \ Assuming that the time-scale for the
fluctuations of the radial mode are sufficiently fast in comparison to the
time-scale\footnote{Whereas in the axion potential discussion we took the
scale of fluctuations for $\theta_{\mathrm{YM}}$ to be sufficiently large
compared to all such glueball phase alignment decay events, here we assume
that the radial mode fluctuates more rapidly. \ As computed in \cite{HSV},
when the size of the glueball field is not exponentially suppressed this decay
rate is an exponentially suppressed quantity so that there are typically a
wide range of energy scales available at which individual modes become
relevant for the analysis.} for the decay of the glueball phase to the
energetically preferred alignment configuration, it follows that there exists
a stable minimum of $V_{\mathrm{rad}}\left(  \Delta\right)  $ for $k,l\sim
N/2$ given by:%
\begin{equation}
\left\vert \Delta_{\ast}\right\vert ^{4}=-20\left\vert \Lambda_{0}\right\vert
^{4}e^{-8\pi^{2}/\lambda}\left(  \cos\frac{2\pi k+\theta_{\mathrm{YM}}}%
{N}+\cos\frac{2\pi l+\theta_{\mathrm{YM}}}{N}\right)  \text{.}%
\label{deltastab}%
\end{equation}
Although highly suggestive, it is unclear whether this mechanism is consistent
with the stabilization of the other moduli of the system. \ Indeed, it follows
from the analysis of \cite{HSV} as well as section \ref{Breakdown} that in the
two cut geometry the size of the glueball fields becomes unstable when
equation (\ref{critequation}) holds so that:%
\begin{equation}
\log\left\vert \frac{\Lambda_{0}}{\Delta}\right\vert ^{4}\sim\frac
{1}{20t_{\ast}}=\frac{1}{20}\left\vert \frac{\Delta}{\Lambda_{0}}\right\vert
^{4}e^{8\pi^{2}/\lambda}\label{breakdownestimate}%
\end{equation}
where in the last equality we have approximated $t_{\ast}$ by its one loop
value. \ This expression is of the same order of magnitude as equation
(\ref{deltastab}). \ Assuming $k,l\sim N/2$, simultaneously satisfying
equation (\ref{deltastab}) and the condition $t<t_{\ast}$ requires:%
\begin{equation}
0<\log\left\vert \frac{\Lambda_{0}}{\Delta}\right\vert ^{4}<2
\end{equation}
which is not consistent with an expansion in large $\Lambda_{0}$. \ Thus, at
the level of analysis presented here, we cannot definitively conclude one way
or the other whether misaligned glueball phases will stabilize the radial mode.

Adding additional $S^{2}$'s does not na\"{\i}vely resolve the difficulties
encountered above. \ For simplicity, we neglect fluctuations corresponding to
the motion of unoccupied minimal size $S^{2}$'s and consider symmetric
configurations of minimal size $S^{2}$'s of the type considered in section
\ref{Breakdown}. \ It follows from equation (\ref{critequation}) that the size
of the glueball field develops an instability at a value comparable to that
given by equation (\ref{breakdownestimate}) where the factor of $20$ is now
replaced by $\left\vert 2\beta\right\vert $, with $\beta$ as in equation
(\ref{betadef}) of appendix B. \ On the other hand, this same replacement will
also occur in equation (\ref{deltastab}).

Before concluding this section, we note that strictly speaking, the analysis
of section \ref{StrongCP} assumes the existence of another mechanism to
stabilize the radial mode. \ This is because the axion potential was obtained
by minimizing over all glueball phases.

\section{Conclusions and Discussion \label{Conclude}}

Changing the location of minimal size $S^{2}$'s triggers phase transitions in
geometrically metastable configurations given by branes and anti-branes
wrapping homologous minimal size $S^{2}$'s in a non-compact Calabi-Yau
threefold. \ While two loop effects appear to generically lift the degeneracy
in vacua present in supersymmetric confining vacua, changing the relative
positions of the $S^{2}$'s causes discrete jumps in which confining vacuum has
minimal energy density. \ The minima of the associated axion potential are
also constrained by the discrete symmetries of a given configuration.
\ Although the presence of additional $S^{2}$'s does not alter the qualitative
conditions for a breakdown in metastability, their presence leads to the
existence of walls of marginal stability for nearly BPS\ brane/anti-brane
annihilation processes. \ This generates novel phase dynamics where branes
\textquotedblleft hop\textquotedblright\ to nearby minimal size $S^{2}$'s.
\ Although far from conclusive, when the glueball phases anti-align with the
energetically preferred vacuum, the two loop correction to the vacuum energy
density may stabilize the radial mode present in any compact Calabi-Yau
threefold geometry. \ In the rest of this section we discuss some implications
of this work and some possible avenues of future investigation.

Over suitable time-scales, the ranks of the low energy gauge groups will
dynamically redistribute due to hopping effects. \ This leads to a statistical
distribution of gauge groups connected by such events. \ It is intriguing that
for geometries with a large number of minimal size $S^{2}$'s, hopping would
seem to favor gauge groups with a large number of low rank gauge group factors.

Along similar lines, the statistical mechanics of hopping may also be of
independent interest. \ In geometries with a large grid of unoccupied minimal
size $S^{2}$'s, one can arrange for a \textquotedblleft
current\textquotedblright\ of branes to flow along the grid. \ It is natural
to speculate on the existence of suitable conditions leading to
superconductivity in this setup. \ In the case of a one dimensional lattice of
minimal size $S^{2}$'s, the presence of hopping also suggests connections with
the physics of spin chains, with the unoccupied $S^{2}$'s playing the role of
impurities. \ It would be interesting to see whether statistical correlations
in such systems undergo phase transitions as one increases the
\textquotedblleft doping\textquotedblright\ of the system.

It may also be important to determine the consequences of hopping from the
perspective of a four dimensional observer. \ As one possibility, note that in
some cases a multi-stage bubble nucleation process will lead to a collision
between distinct bubbles. \ While the collision of such bubbles of vacuum has
been studied from various perspectives, it would be interesting to see how the
results of this paper fit with these other developments.

Walls of marginal stability for brane/anti-brane annihilation processes are
most likely a generic feature of compact Calabi-Yau threefolds, although there
are likely to be further features specific to the compact case. \ For example,
the corresponding domain wall solutions may not factorize as the product of a
path on a local Riemann surface with an $S^{2}$ fibration. \ Determining the
\textquotedblleft mean free path\textquotedblright\ for such obstructions in a
compact Calabi-Yau threefold would be interesting.

There are many additional subtleties present for brane/anti-brane systems on
compact Calabi-Yau threefolds. \ In addition to $\alpha^{\prime}$ corrections
to the form of the K\"{a}hler metric, one must also find a mechanism which
stabilizes all of the non-normalizable modes of the non-compact case. \ It is
very suggestive that glueball fields with phases anti-aligned with the
energetically preferred confining vacuum generate a power law repulsion term
which can na\"{\i}vely stabilize the radial mode of a brane/anti-brane system.
\ Because this result is highly contingent on the phase of the glueball
fields, in the absence of another mechanism, stabilization of the radial mode
would appear to preferentially select metastable confining vacua with
\textit{maximal} rather than minimal energy density. \ Note that this is
somewhat at odds with the procedure outlined in section \ref{StrongCP} for
extracting the axion potential by minimizing over all glueball phases. \ It is
likely that further phase structure results from the interplay between the
stabilization of the axion, the radial mode, and the other closed string modes
of compact Calabi-Yau threefolds.

\section*{Acknowledgements}

We thank J. Seo for many helpful discussions and collaboration at an early
stage of this work. \ In addition, we thank M. Aganagic, C. Beasley, M.R.
Douglas, A.L. Fitzpatrick, S. Kachru, J. Marsano, K. Papadodimas, M. Shigemori
and L.-S. Tseng for helpful discussions. \ CV thanks the CTP\ at MIT for
hospitality during his sabbatical leave, where part of this work was
completed. \ The work of the authors is supported in part by NSF grants
PHY-0244821 and DMS-0244464. \ The research of JJH\ is also supported by an
NSF\ Graduate Fellowship.

\appendix

\section*{Appendix A: Two Loop Corrections to $\tau_{ij}$}

In this appendix we collect the two loop corrections to $\tau_{ij}$ derived in
section \ref{MatrixModel}. \ For $i\neq j$ we have:%
\begin{align}
2\pi i\tau_{ij}  &  =-\log\frac{\Lambda_{0}^{2}}{\Delta_{ij}^{2}}+\frac
{1}{W^{(2)}\left(  a_{i}\right)  }\frac{2}{\Delta_{ij}^{2}}\left(
-1-\Delta_{ij}\frac{W^{(3)}\left(  a_{i}\right)  }{W^{(2)}\left(
a_{i}\right)  }+2\frac{W^{(2)}(a_{i})}{W^{(2)}(a_{j})}\right)  S_{i}\\
&  +\frac{1}{W^{(2)}\left(  a_{j}\right)  }\frac{2}{\Delta_{ji}^{2}}\left(
-1-\Delta_{ji}\frac{W^{(3)}\left(  a_{j}\right)  }{W^{(2)}\left(
a_{j}\right)  }+2\frac{W^{(2)}(a_{j})}{W^{(2)}(a_{i})}\right)  S_{j}\\
&  +\underset{k\notin\left\{  i,j\right\}  }{\sum}\left(  \frac{4}%
{W^{(2)}(a_{i})\Delta_{ij}\Delta_{ik}}+\frac{4}{W^{(2)}(a_{j})\Delta
_{ji}\Delta_{jk}}+\frac{4}{W^{(2)}(a_{k})\Delta_{ki}\Delta_{kj}}\right)
S_{k}+\mathcal{O}(S^{2})\text{ }%
\end{align}
and, for terms on the diagonal:%
\begin{align*}
2\pi i\tau_{ii}  &  =\log\frac{S_{i}}{W^{(2)}(a_{i})\Lambda_{0}^{2}}+\left(
\frac{W^{(3)}\left(  a_{i}\right)  ^{2}}{W^{(2)}(a_{i})^{3}}-\frac{1}{2}%
\frac{W^{(4)}(a_{i})}{W^{(2)}(a_{i})^{2}}\right)  S_{i}\\
&  +\underset{i\neq j}{%
{\displaystyle\sum}
}\frac{1}{W^{(2)}\left(  a_{i}\right)  }\frac{2}{\Delta_{ij}^{2}}\left(
-1-\Delta_{ij}\frac{W^{(3)}\left(  a_{i}\right)  }{W^{(2)}\left(
a_{i}\right)  }+2\frac{W^{(2)}(a_{i})}{W^{(2)}(a_{j})}\right)  S_{j}%
+\mathcal{O}(S^{2})\text{.}%
\end{align*}

As a check on the above computations, we next compare our value of the two
loop correction to $\tau$ which we denote by $\delta\tau$ with the known
values for the two and three cut geometries. \ When $W^{\prime}(x)=g(x-a_{1}%
)(x-a_{2})$, this reduces to:%
\begin{align}
2\pi i\delta\tau_{11}^{2-\mathrm{cut}}  &  =4\frac{S_{1}}{g\Delta_{12}^{3}%
}-10\frac{S_{2}}{g\Delta_{12}^{3}}\\
2\pi i\delta\tau_{12}^{2-\mathrm{cut}}  &  =-10\frac{S_{1}}{g\Delta_{12}^{3}%
}+10\frac{S_{2}}{g\Delta_{12}^{3}}%
\end{align}
which agrees with appendix B of \cite{CIV}.

When $W^{\prime}(x)=g(x-a_{1})(x-a_{2})(x-a_{3})$, we find\footnote{We thank
J. Seo for suggesting this check.}:%
\begin{align}
2\pi i\delta\tau_{11}^{3-\mathrm{cut}}  &  =\frac{1}{g\Delta_{12}^{2}%
\Delta_{13}^{2}}\left(  5+4\frac{\Delta_{13}}{\Delta_{12}}+4\frac{\Delta_{12}%
}{\Delta_{13}}\right)  S_{1}\\
&  +\frac{2}{g\Delta_{12}^{2}\Delta_{13}\Delta_{23}}\left(  -2+5\frac
{\Delta_{23}}{\Delta_{21}}-2\frac{\Delta_{32}}{\Delta_{31}}\right)  S_{2}\\
&  +\frac{2}{g\Delta_{13}^{2}\Delta_{12}\Delta_{32}}\left(  -2+5\frac
{\Delta_{32}}{\Delta_{31}}-2\frac{\Delta_{23}}{\Delta_{21}}\right)  S_{3}\\
2\pi i\delta\tau_{12}^{3-\mathrm{cut}}  &  =\frac{2}{g\Delta_{12}^{2}%
\Delta_{13}\Delta_{23}}\left(  -2+5\frac{\Delta_{23}}{\Delta_{21}}%
-2\frac{\Delta_{32}}{\Delta_{31}}\right)  S_{1}\\
&  -\frac{2}{g\Delta_{21}^{2}\Delta_{23}^{2}}\left(  2+2\frac{\Delta_{32}%
}{\Delta_{31}}+5\frac{\Delta_{23}}{\Delta_{21}}\right)  S_{2}\\
&  +\frac{8}{g\Delta_{12}\Delta_{13}\Delta_{23}^{2}}\left(  1-\frac
{\Delta_{23}}{\Delta_{21}}-\frac{\Delta_{32}}{\Delta_{31}}\right)  S_{3}%
\end{align}
which agrees\footnote{In appendix C of \cite{CVNOT}, the sign of what is
referred to as $h_{ab}$ should be reversed in the period $\Pi_{a}$. \ This is
easily checked (and remedied) by appealing to the symmetry of the $\tau$
matrix entries under permutation of its indices.} with appendix C of
\cite{CVNOT}.

\section*{Appendix B: Two Cut $\tau_{ij}$ with $n$ Minimal Size $S^{2}$'s}

In this appendix we compute the form of the period matrix when:%
\begin{equation}
W^{\prime}(x)=g(x-a_{1})(x-a_{2})f(x)
\end{equation}
where $f$ is a polynomial with isolated roots and all $A$-cycle periods other
than $S_{1}$ and $S_{2}$ are formally set to zero, in accord with the
discussion in section \ref{Unoccupied}.

Setting $t_{i}\equiv S_{i}/(W^{(2)}(a_{i})\Delta_{12}^{2})$, it follows from
the expressions given in appendix A that the relevant matrix entries are:%
\begin{align}
2\pi i\tau_{11}  &  =\log\left(  t_{1}\frac{\Delta_{12}^{2}}{\Lambda_{0}^{2}%
}\right)  +\left[  \left(  2+\frac{2\Delta_{12}f^{\prime}(a_{1})}{f(a_{1}%
)}\right)  ^{2}-\frac{\Delta_{12}}{2}\frac{6f^{\prime}(a_{1})+3\Delta
_{12}f^{(2)}(a_{1})}{f(a_{1})}\right]  t_{1}\\
&  +\left[  4+2\frac{f(a_{2})}{f(a_{1})}+4\frac{f(a_{2})}{f(a_{1})}\left(
1+\frac{\Delta_{12}f^{\prime}(a_{1})}{f(a_{1})}\right)  \right]
t_{2}+\mathcal{O}(t^{2})\\
2\pi i\tau_{12}  &  =-\log\frac{\Lambda_{0}^{2}}{\Delta_{12}^{2}}+\left[
-6-4\frac{f(a_{1})}{f(a_{2})}-4\frac{\Delta_{12}f^{\prime}(a_{1})}{f(a_{1}%
)}\right]  t_{1}\\
&  \left[  -6-4\frac{f(a_{2})}{f(a_{1})}+4\frac{\Delta_{12}f^{\prime}(a_{2}%
)}{f(a_{2})}\right]  t_{2}+\mathcal{O}(t^{2})\\
2\pi i\tau_{22}  &  =\log\left(  t_{2}\frac{\Delta_{12}^{2}}{\Lambda_{0}^{2}%
}\right)  +\left[  \left(  2-\frac{2\Delta_{12}f^{\prime}(a_{2})}{f(a_{2}%
)}\right)  ^{2}+\frac{\Delta_{12}}{2}\frac{6f^{\prime}(a_{2})-3\Delta
_{12}f^{(2)}(a_{2})}{f(a_{2})}\right]  t_{2}\\
&  +\left[  4+2\frac{f(a_{1})}{f(a_{2})}+4\frac{f(a_{1})}{f(a_{2})}\left(
1-\frac{\Delta_{12}f^{\prime}(a_{2})}{f(a_{2})}\right)  \right]
t_{1}+\mathcal{O}(t^{2})\text{.}%
\end{align}
When $f(x)$ is an even or odd polynomial whose roots are all isolated and
real, the entries of $\tau_{ij}$ and $\mathcal{F}_{ijk}$ simplify further:%
\begin{align}
\left[
\begin{array}
[c]{cc}%
2\pi i\tau_{11} & 2\pi i\tau_{12}\\
2\pi i\tau_{12} & 2\pi i\tau_{22}%
\end{array}
\right]   &  =\left[
\begin{array}
[c]{cc}%
\log\left(  t_{1}\frac{\Delta_{12}^{2}}{\Lambda_{0}^{2}}\right)  +\gamma
t_{1}+\beta t_{2} & -\log\frac{\Lambda_{0}^{2}}{\Delta_{12}^{2}}-(-1)^{\deg
f}\beta\left(  t_{1}+t_{2}\right) \\
-\log\frac{\Lambda_{0}^{2}}{\Delta_{12}^{2}}-(-1)^{\deg f}\beta\left(
t_{1}+t_{2}\right)  & \log\left(  t_{2}\frac{\Delta_{12}^{2}}{\Lambda_{0}^{2}%
}\right)  +\gamma t_{2}+\beta t_{1}%
\end{array}
\right] \\
\left[
\begin{array}
[c]{cc}%
2\pi i\mathcal{F}_{111} & 2\pi i\mathcal{F}_{112}\\
2\pi i\mathcal{F}_{122} & 2\pi i\mathcal{F}_{222}%
\end{array}
\right]   &  =\left[
\begin{array}
[c]{cc}%
\frac{1}{g\Delta_{12}^{3}f(a_{1})}\left(  \frac{1}{t_{1}}+\gamma\right)  &
\frac{-\beta}{g\Delta_{12}^{3}f(a_{2})}\\
\frac{\beta}{g\Delta_{12}^{3}f(a_{1})} & \frac{-1}{g\Delta_{12}^{3}f(a_{2}%
)}\left(  \frac{1}{t_{2}}+\gamma\right)
\end{array}
\right]
\end{align}
where we have introduced the parameters:%
\begin{align}
\gamma &  =\left(  2+\frac{2\Delta_{12}f^{\prime}(a_{1})}{f(a_{1})}\right)
^{2}-\Delta_{12}\frac{6f^{\prime}(a_{1})+3\Delta_{12}f^{(2)}(a_{1})}%
{2f(a_{1})}\label{gammadef}\\
\beta &  =4+6(-1)^{\deg f}+4(-1)^{\deg f}\frac{\Delta_{12}f^{\prime}(a_{1}%
)}{f(a_{1})}\text{.} \label{betadef}%
\end{align}
\newpage
\bibliographystyle{ssg}
\bibliography{multicut}

\end{document}